\documentclass[prb,
twocolumn,
nopacs,
nofootinbib,
superscriptaddress]{revtex4-2}

\usepackage{bm}
\usepackage{physics}
\usepackage{graphicx}
\usepackage{subcaption} 
\usepackage{booktabs}
\usepackage{amsmath, amssymb}

\usepackage[colorlinks,
    linkcolor=blue,
    anchorcolor=red,
    citecolor=red
]{hyperref}
\usepackage{wasysym}
\usepackage{MnSymbol}
\usepackage{comment}
\usepackage[dvipsnames]{xcolor}
\usepackage[normalem]{ulem}\usepackage[normalem]{ulem}

\usepackage{wasysym}
\usepackage{caption}
\usepackage{ragged2e}

\renewcommand{\v}[1]{\mathbf{#1}}

\begin{document}
\global\long\def\P{\mathcal{P}}%
\global\long\def\u{\upsilon}%

\title{Anomalous Hall effect in rhombohedral graphene}

\author{Vera Mikheeva}
\affiliation{Nanocenter CENN, Ljubljana, Slovenia}
\affiliation{Faculty of Mathematics and Physics, University of Ljubljana, Ljubljana, Slovenia}

\author{Daniele Guerci}
\affiliation{Department of Physics, Massachusetts Institute of Technology, Cambridge, USA}

\author{Daniel Kaplan}
\email{d.kaplan1@rutgers.edu}
\affiliation{Department of Physics and Astronomy, Center for Materials Theory, Rutgers University, Piscataway, USA}

\author{Elio J. K\"onig}
\affiliation{Department of Physics, University of Wisconsin-Madison, Madison, Wisconsin, USA}
\affiliation{Max-Planck Institute for Solid State Research, Stuttgart, Germany}

\begin{abstract}
    Motivated by recent experiments on rhombohedral stacked multilayer
graphene and the observation of the anomalous Hall effect in a spontaneous spin-valley polarized quarter metal state, we calculate the anomalous Hall
conductivity for this system in the presence of two types of impurities: weak and dense
as well as sparse
and strong. 
Our calculation of $\sigma_{xy}$ is based on the Kubo-Streda diagrammatic approach.  In a model with Gaussian disorder applicable to weak dense impurities, this involves all non-crossing diagrams (intrinsic, side-jump and Gaussian skew-scattering contributions) and additionally diagrams with two intersecting impurities, X and $\Psi$, representing diffractive skew-scattering processes. 
A "Mercedes star" diagram (non-Gaussian skew scattering) is furthermore included to treat in
the case of strong, sparse impurities. 
We supplement our asymptotically exact analytical solutions for an isotropic model without warping effects by semi-numerical calculations accounting perturbatively for
warping, which plays a crucial role in the low-energy band structure. 
\end{abstract}

\maketitle

Graphene \cite{novoselov2005two,geim2007rise,shi2020electronic} based structures are an active playground for interacting and correlated phases in condensed matter physics. Experimental and theoretical interest in these materials has particularly sparked since the recent discovery of
a rich spectrum of correlated
electron phenomena, including superconductivity, correlated insulators, spontaneous spin and valley polarization, anomalous Hall crystals, and fractional quantum anomalous Hall states \cite{han2024correlated, han2025signatures, zhou2021half, han2024engineering, geisenhof2021quantum, pantaleon2023superconductivity, lu2025extended, zhou2020enhanced, zhou2022disorder, seiler2024probing, choi2025superconductivity, yang2025impact, mu2025valley,  cao2018unconventional,zhou2021superconductivity, lu2024fractional, yankowitz2019tuning, liu2021tuning,Martinez2025}.
Since most of the correlated phases are found in the rhombohedral ($ABC$) stacked configuration,
here we focus on stacking $n$-layer graphene, a metastable but experimentally realizable \cite{zhang2025layer, kumar2025superconductivity} 
configuration for multilayer graphene (for $n>2$, $ABA$ stacking is energetically advantageous). 
Unlike moir\'e systems, they exhibit a plethora of phases without breaking the atomic crystal's translation symmetry and still display an increased electronic density of states (DOS) near charge neutrality \cite{Ghazaryan2023}. This remarkable property is due to the structure of low-energy states permitted by the rhombohedral arrangement of layers \cite{cao2024rhombohedral}: the low energy states are localized on opposite sublattices of the extremal layers ($A_1,B_N$, where $N$ is the number of layers), with the bulk states gapped out by hybridization. 
 
Thus,
the low energy bands flatten with $N$ and are further modified by the application of a nonzero perpendicular displacement field separating the states of the $(A_1, B_N)$ pseudospinor, resulting in a tunable gap at the $K$/$K'$ points. The sizable valley-dependent Berry curvature inherited from graphene monolayers furthermore allows that both correlated and topological states can emerge from this
material, including the 
spin-polarized half-metal and spin- and valley-polarized quarter-metal phases 
which effectively result from a Stoner-like instability. These phases spontaneously break time reversal symmetry and, amongst others,  
were observed  
by measurements of the anomalous Hall effect (AHE) \cite{zhou2021half, han2024correlated, choi2025superconductivity}. 
\begin{figure}[h!]
    \centering

    \begin{minipage}[t]{0.5\linewidth}
        \centering
        \includegraphics[width=\linewidth]{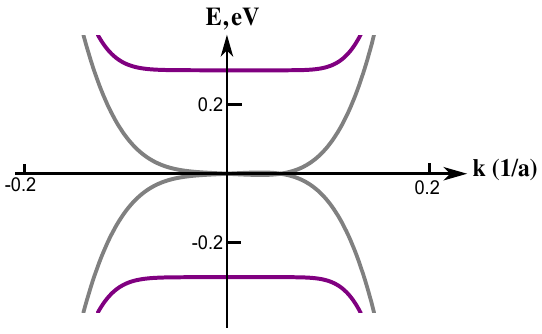}
        \caption*{(a)}
    \end{minipage}%
    \hfill
    \begin{minipage}[t]{0.5\linewidth}
        \centering
        \includegraphics[width=\linewidth]{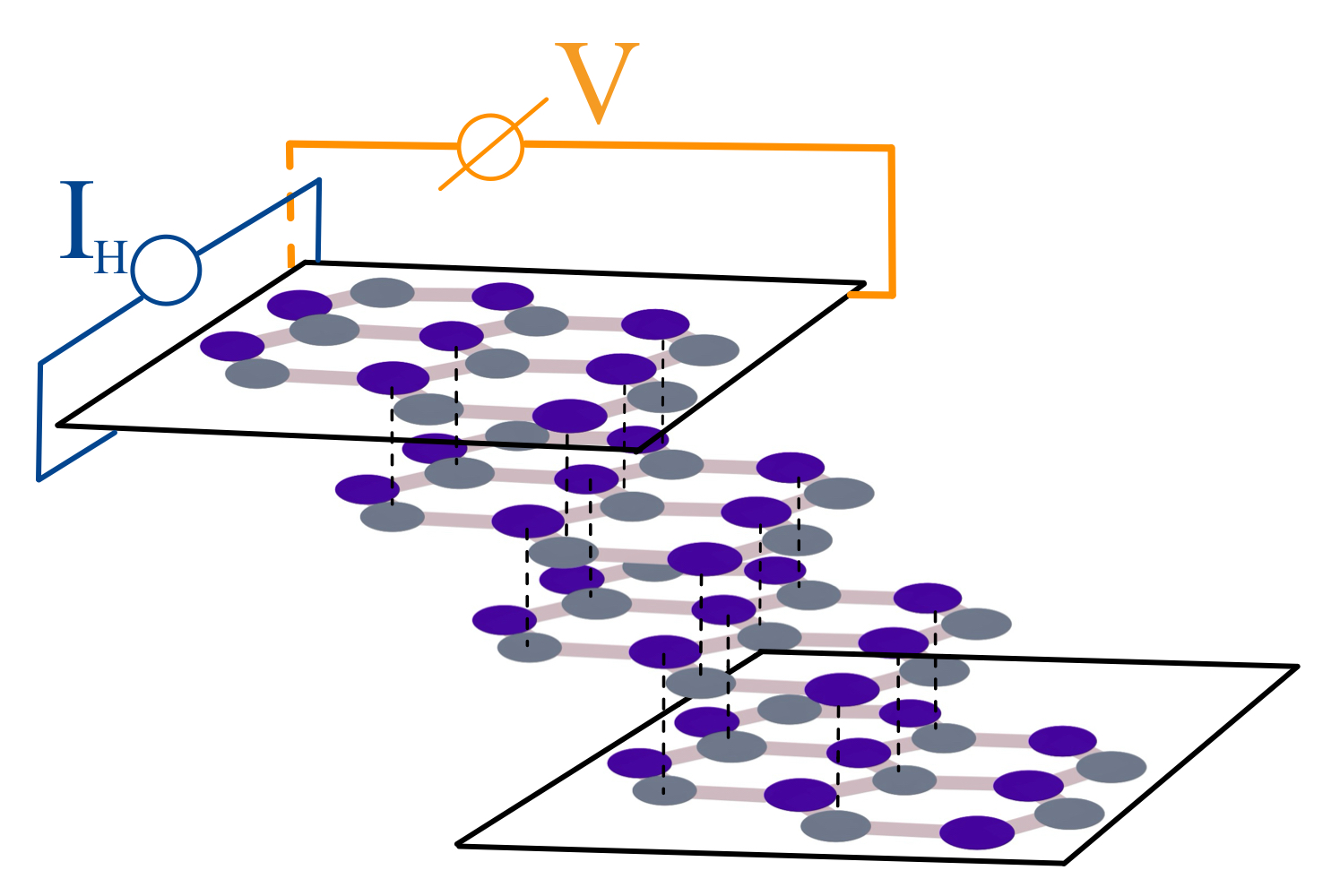}
        \caption*{(b)}
    \end{minipage}

    \vspace{0.4em}

   \captionsetup{justification=justified, singlelinecheck=false}

    \caption{\justifying (a) Schematic band structure of rhombohedral multilayer  
    graphene without (grey) and with displacement field $m=0.3$ eV (purple). Here $a=0.246$ nm is the graphene lattice constant; (b) Schematic system and 
    experimental setup for measuring the anomalous Hall effect in tetralayer rhombohedral graphene.}
    \label{fig: combo}
\end{figure}

The AHE, i.e.~transverse voltage in response to a current, without applying an external magnetic field, \cite{nagaosa2010anomalous} was discovered in 1881, by E. Hall in a study of ferromagnets.  
 For this effect to exist, the system must  
 break 
 time reversal symmetry and display a non-trivial chiral texture in momentum space  
 transforming the information of  
 spin or valley polarization to orbital motion. Rhombohedral graphene satisfies these demands in the quarter-metal phase. Multiple physical effects conspire in a non-zero AHE signal. The intrinsic contribution \cite{karplus1954hall, jungwirth2002anomalous} is determined solely by the electronic structure of the material, specifically the Berry curvature of occupied states. 
Additionally, extrinsic effects are of relevance. Amongst those, impurity scattering is most prominent at lowest temperatures and its contribution is parametrically the same or  
can even become the leading source of anomalous Hall response in realistic systems, where disorder is always present to some extent. 

 To describe those phenomena,  
 there are two common techniques based on Boltzmann kinetic equation  \cite{sinitsyn2007semiclassical,  nagaosa2010anomalous} and Kubo-Streda diagrammatic formalism \cite{streda1982theory, nunner2007anomalous}. 
In this paper, we will focus on the latter approach. However, we  
note that both techniques are 
equivalent and both commonly used. 

Theoretically,  
the impurity induced contributions to the total anomalous Hall conductivity $\sigma_{xy}$ can be decomposed into two distinct contributions associated with different physical mechanisms:  
side-jump 
\cite{berger1964influence,belinicher1982kinetic,sinitsyn2006coordinate} and skew-scattering \cite{smit1955spontaneous}. 
The side-jump mechanism arises due to a transverse displacement of the electron wave packet during impurity scattering i.e., 
semiclassical trajectories of incoming and outgoing electrons do intersect away from the scattering center.
The skew-scattering mechanism, resulting from an asymmetry (broken detailed balance) in the impurity scattering cross section, is conveniently divided into three distinct components: Gaussian skew scattering
 (non-crossing diagrams) \cite{sinitsyn2007anomalous}, diffractive (crossed diagrams) \cite{ado2015anomalous, ado2016anomalous, konig2016anomalous}, and non-Gaussian (higher moments of the random potential) \cite{nagaosa2010anomalous,lutchyn2009frequency, goryo2008impurity}. Some authors reserve the word "skew-scattering" to the non-Gaussian effect ("skewness" of the disorder distribution function) but we employ the more general definition here. The Gaussian and diffractive contributions to skew scattering stem from complexes of nearby impurities and are therefore parametrically comparable. Gaussian skew scattering stems from coherent interband transitions. 
In contrast, diffractive skew scattering is a quantum interference effect.  

In this paper, we consider the anomalous Hall effect in rhombohedral multilayer graphene in the presence of disorder using the diagrammatic Kubo-Streda approach. As a disorder, we consider two different cases: Gaussian (weak dense impurities) and non-Gaussian disorder (strong sparse impurities) 
which is modeled phenomenologically via a finite third moment of the disorder potential.
For weak dense impurities we calculate the response to leading order in disorder strength, i.e.~we take into account the ladder diagrams and diagrams with two intersecting impurity lines. We present the general analytical result for intrinsic and non-crossing contributions and the general answer including diffractive skew-scattering. Besides the isotropic model, we consider  
trigonal warping \cite{ando1998berry, koshino2009trigonal, koshino2010interlayer, nagaosa2010anomalous} (an effect of the crystal lattice), which plays a crucial role in the low-energy band structure. We demonstrate using numerical calculations how it affects the anomalous Hall conductivity. Throughout, we work in a single-valley  theory and neglect intervalley scattering, which requires momentum transfer $|\mathbf K-\mathbf K'|$ -- it is thus suppressed for smooth disorder -- and is moreover off-resonent in the quarter metal phase. 
  In the following, we consider the general case of rhombohedral multilayer graphene with an arbitrary number of layers $n$. However, we focus on the cases $n = 2, 3, 4, 5$  for isotropic and $n = 3,4$ for warped model,  which are the most relevant from the experimental point of view.

This paper is structured as follows: in the next section (Sec. \ref{sec: model}), we describe the effective low-energy Hamiltonian of ABC-stacked multilayer graphene and discuss disorder types with respect to experimental reality.
Then, in Sec.~\ref{sec: isotropic}, we consider an isotropic model and, using diagrams, we present the full result for $\sigma_{xy}$. In Sec.~\ref{sec: warped}, we provide the same calculations in the 
presence of interlayer asymmetry
that opens a gap in the spectrum. Section~\ref{sec: comparison} dedicates the
comparison between isotropic and warped models with numerical
calculations.

\section{Model}
\label{sec: model}

The  
low-energy electronic properties of rhombohedral graphene are described by 
an effective
 two-component Hamiltonian that describes 
wavefunctions predominantly localized on atomic sites $A$ of the bottom layer and $B$ of the top layer
\cite{koshino2009trigonal, guinea2006electronic, lu2006absorption} (Appendix~\ref{ap: Hamiltonian}):
\begin{equation}
    H_n^{(w)} =  \left (\begin{array}{cc}
       m & v\mathcal P ^n + w\mathcal P^{n - 3} \\
      v \overline{\mathcal P} ^n + w\overline{\mathcal P}^{n - 3} &-m
   \end{array} \right) = \vec d(\v p) \cdot \vec \sigma,
   \label{eq: effH}
\end{equation} 
where $\P = \xi p_x - ip_y$ and $\xi = \pm 1$ is the valley index.
We focus on a single valley and 
will keep $\xi =1$; $v$ is a constant with dimensions [energy × length$^n$]; $\vec \sigma$ is the vector of Pauli matrices describing sublattice space;  $m$ is the displacement field; the term proportional to $w$ describes trigonal warping. 
The $n$-th power $\P^n$ in the momentum
  dependence of the effective low-energy Hamiltonian originates from the virtual interlayer hopping processes connecting the sublattices in $n$-layer rhombohedral stack.
Trigonal warping arises from subleading interlayer hoppings, which couple equivalent sublattices across adjacent layers and break continuous rotational symmetry. 
We note that the effective Hamiltonian Eq.~\eqref{eq: effH} describes multilayer graphene in rhombohedral graphene for $n \geq 3$ (the low energy limit of Bernal bilayer graphene is also captured by taking $n = 2$ and $w = 0$).

Impurities are modeled by a random potential $V(\mathbf x)$ with only two non-vanishing moments
\begin{subequations}
\begin{align}
     \langle V(\mathbf{x}_1) V(\mathbf{x}_2) \rangle &=2 \pi \alpha v^2 p_0 ^{2n-2} \delta (\mathbf{x}_{1} - 
     \mathbf{x}_{2}) ,
     \label{eq: gaus}\\
    \langle V(\mathbf{x}_1) V(\mathbf{x}_2) V(\mathbf{x}_3) \rangle  &=\beta v^3 p_0 ^{3n-4} \delta(\mathbf{x}_1 - \mathbf{x}_2) \delta(\mathbf{x}_2 - \mathbf{x}_3) ,
    \label{eq: nongaus}
\end{align}
\label{eq: disorder}
\end{subequations}
where $\alpha$ and $\beta$ are a
dimensionless parameter characterizing disorder strength \cite{ado2015anomalous, lutchyn2009frequency}. 
We introduced the Fermi momentum in the absence of warping 
\begin{align}
    p_0 = [(\epsilon_F ^2 - m^2)/v^2]^{\frac{1}{2n}},
\end{align}
where $\epsilon_F$ is the Fermi energy.

We briefly comment on the experimental relevance of our model of impurity scattering in rhombohedral graphene. First, our model explicitly assumes that the dominant scattering mechanism comes from the short-range disorder. The applicability of delta-function impurity potentials in rhombohedral graphene arises from the fact that the characteristic distance between impurities and the gate $(d \sim nm)$ is significantly smaller than the Fermi wavelength $(\lambda_F \sim 100 nm)$. This condition ensures that the impurity potential is screened on a length scale much shorter than that of the electronic wave functions, thus supporting the use of a point-like model. 
Second, we assume that disorder is relevant for transport in our model. 
To verify this, one can compare the mean free path with the sample size. 
The mean free path is estimated to be in the range $l \sim 0.1\text{--}1\ \mu\mathrm{m}$~\cite{zhou2021superconductivity, zhou2021half}, while in experiments the typical sample size is $L \sim 2.5\ \mu\mathrm{m}$~\cite{han2024correlated, zhou2021superconductivity}. 
Thus, samples can be expected slightly on the diffusive side of the ballistic-to-diffusive crossover.

 We employ the diagrammatic approach to calculate the anomalous Hall conductivity of rhombohedral multilayer graphene. In the limit $\beta = 0$ we take 
into account all the diagrams to the leading order in $(p_0 l)^0 \sim \alpha^0$. Note that this involves a non-perturbative resummation of infinitely many diagrams. In addition, we add (perturbatively) the contribution of $\beta$. 

 Technically, the Hall conductivity $\sigma_{xy}$ using the Kubo-Streda formula \cite{sinitsyn2006charge} can be written as $\sigma_{xy}=\sigma_{xy}^{\textsc{I}} + \sigma_{xy}^{\textsc{II}}$ \cite{ado2015anomalous} where 
\begin{subequations}
\begin{align}
    &\sigma_{xy}^{\textsc{I}}=  \frac{e^2}{h} \left\langle \mathrm{Tr} \left[ \hat{j}_x G^R \hat{j}_y G^A \right] \right\rangle,  \\
&\sigma_{xy}^{\textsc{II}} = e c \frac{\partial q}{\partial B}.
\label{eq: congen}
\end{align}
\end{subequations}
Here, 
${ G^{R/A}}$
are the retarded and advanced Green's functions; $j_{x/y} (p)$ are current operators which can be determined as $j_{x/y}(p) = \frac{\partial H_n}{\partial p_{x/y} }$; $q$ is the total electron concentration; $B$ is the external magnetic field; $c$ is the speed of light in vacuum. In this paper we set $\hbar = 1$ unless explicitly highlighted otherwise.

There are two different regimes for Hall conductivity: when the Fermi energy $\epsilon_F$ lies in the gap, $|\epsilon_F|< |m|$ and otherwise. In the former case, the term $\sigma_{xy}^{\textsc{II}}$ in Eq.~\eqref{eq: congen} provides the dominant topologically quantized contribution and can be interpreted in terms of the Chern number.
Outside the gap, i.e., for $|\epsilon_F| > |m|$, this topological contribution becomes negligible \cite{jungwirth2002anomalous,ado2015anomalous}, and the entire Hall conductivity stems from $\sigma_{xy}^{\rm I}$ (diagrammatically a bubble).  Most of our calculations will be in the latter regime. The assumption $p_0 l \gg 1$ ceases to be correct near the band edge. Sufficiently below the band edge (inside the gap) corrections to the topological quantized response are negligible. We note that we entirely neglect Anderson localization physics, e.g. weak-localization corrections  \cite{mitra2007weak, konig2014half}. This is justified for the present case of sufficiently small samples or when dephasing is strong enough.

\begin{figure}
    \centering
    \includegraphics[width=1\linewidth]{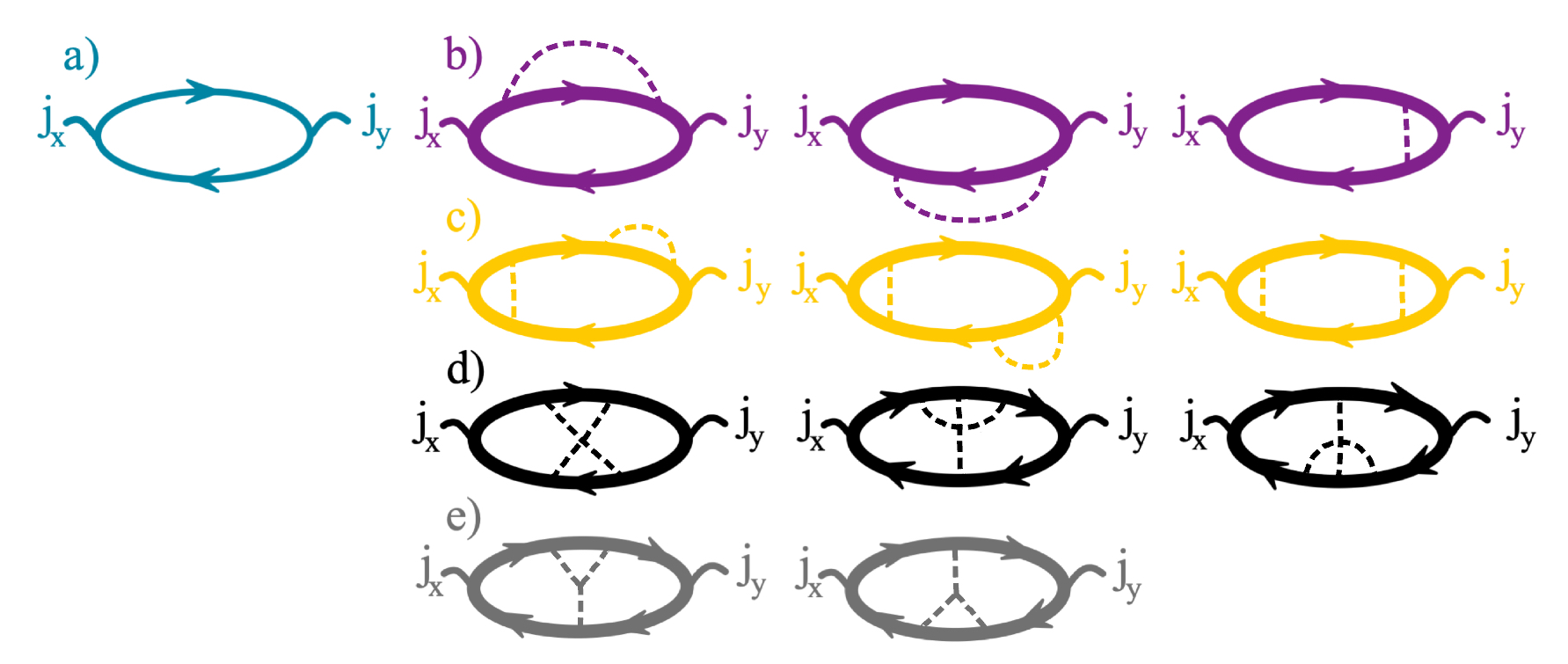}
    \caption{\justifying Exemplary diagrams for the anomalous Hall conductivity: a) intrinsic contribution, b) side jump, c) Gaussian  
    skew-scattering, d) diffractive skew-scattering, e) third moment (skewness).}
    \label{fig: alldiagrams}
\end{figure}

We first focus on $\beta = 0$.   The leading-order contribution in terms of $(p_0 l)$ involves summing ladder diagrams and including the leading correction from crossed impurity lines, together with self-energy corrections required for a conserving approximation (self-consistent Born approximation supplemented with one crossing). 
In our calculation the conductivity bubble is formulated with disorder-averaged Green's functions and renormalized vertices throughout; Fig.~\ref{fig: alldiagrams} serves as a schematic classification of distinct mechanisms (intrinsic, side-jump, and skew scattering) rather than indicating additional self-energy dressings.
Accordingly, thin lines in Fig.~\ref{fig: alldiagrams} should be understood as a schematic notation for bare propagators, while bold lines denote disorder-averaged (dressed) Green’s functions, and dashed lines represent disorder correlators, Eq.~\eqref{eq: disorder}. We notice, that a nonzero $\sigma_{xy}$ requires at least one off-shell Green’s-function segment, which naturally groups two nearby impurities into an effective scattering complex. Once this close pair is treated as a single object, the diagram’s leading order (in $\alpha$, equivalently in $1/(p_0 l)$) is determined by the remaining independent crossings; therefore, additional impurity lines do not automatically imply a higher order in $\alpha$. 

\section{Isotropic model}
\label{sec: isotropic}
In this section we present calculations for
the isotropic model, i.e.~setting $w = 0$ in Eq.~\eqref{eq: effH}. 

\subsection{Intrinsic contribution}

The intrinsic contribution (bare bubble without impurity lines, Fig.~\ref{fig: alldiagrams}, a)) yields:
\begin{align}
    &\sigma_{xy} ^{\text{int}} = - \frac{e^2}{h} \frac{m }{2\epsilon_F} n;  \; |\epsilon_F|> |m|.
    \label{eq: int}
\end{align}
The half-integer quantization if the Fermi energy is inside the gap arises from fermion number fractionalization \cite{jackiw1976solitons}-\cite{konig2014half}. (cf.~Appendix~\ref{ap: int_isotropic}). As a technical remark to obtain a non-zero Hall conductivity, we note that it is essential that one of the Green's functions in the diagram lies away from the Fermi energy (i.e.,
is off-shell).

\subsection{Non-Crossing approximation}

Next, we study
how the presence of impurities influences the anomalous Hall conductivity. We can use the fact that the self-consistent Born approximation supplemented by diagrams with one crossed line yields a self-energy that is essentially the same as in the standard Born approximation, allowing us to obtain the average retarded Green's function: 
\begin{align}
    \mathbf{G}_{\mathbf{p}}^R  =\frac{\epsilon_F \left (1 + \frac{\pi i \alpha  }{2n} \right ) + m\sigma_z \left (1 - \frac{\pi i \alpha  }{2n} \right ) +d_x(\v p) \sigma_x +d_y (\v p) \sigma_y}{\epsilon_F ^2 - m^2 - d_x(\v p)^2 - d_y(\v p)^2 + \frac{i \pi \alpha }{n } (\epsilon_F ^2 +m^2)}.
    \label{eq: momentumGF}
\end{align}
The advanced Green's function is the Hermitian conjugate of
the above expression (Appendix~\ref{ap: SCBA}).

\begin{figure}
    \centering
    \includegraphics[width=1\linewidth]{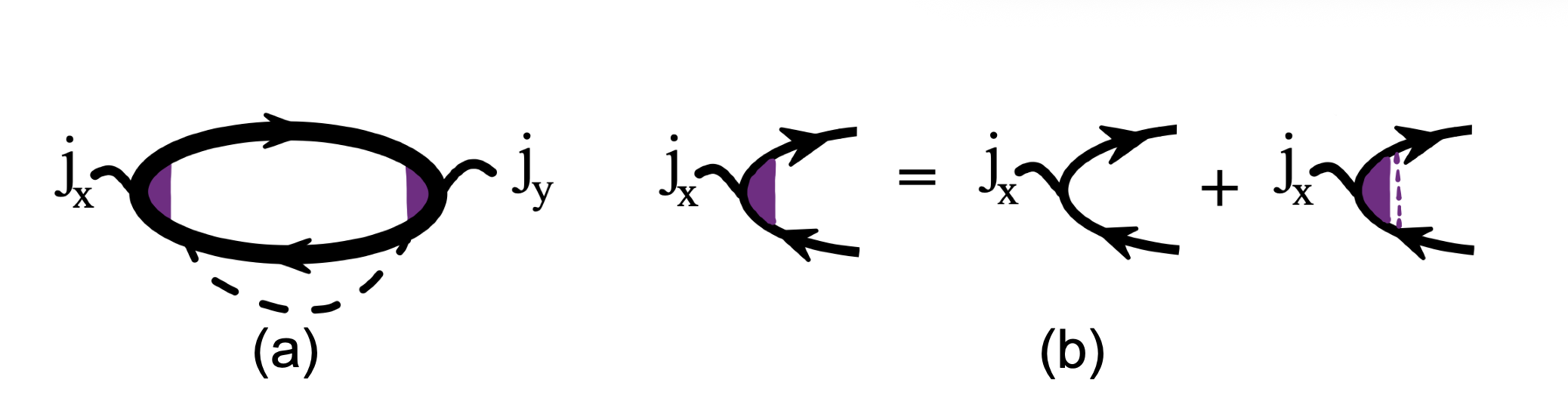}
    \caption{\justifying Example diagram of the side-jump contribution with a vertex correction (a), which is obtained by summing ladder diagrams (b).}
    \label{fig: vertex}
\end{figure}

We use this in the evaluation of 
side-jump and skew-scattering terms. First, we focus on non-crossing contribution. It is worth noting that diagrams with the oﬀ-shell Green function, which links the off-diagonal current vertex to an impurity, correspond to the side-jump contribution, while in skew-scattering diagrams, the off-shell Green's function connects two impurities. 

As the
first order correction, Fig.~\ref{fig: vertex}, to the vertex
\begin{align}
    \delta j_{x/y} \propto  
   \alpha \int \frac{d^2 p}{(2\pi)^2} \mathbf{G}_{\mathbf{p}} ^A  j_{x/y}  \mathbf{G}_{\mathbf{p}} ^R = 0; \; n \neq 1,
    \label{eq: correction_isotropic}
\end{align}
vanishes due to the momentum structure in $j_{x/y}= \frac{\partial H}{\partial p_{x/y}}$, bare and dressed vertices coincide.
We emphasize that the vertex correction is non-zero for $n=1$ \cite{ado2015anomalous,nagaosa2010anomalous, sinitsyn2007anomalous}. The contribution of Gaussian
skew-scattering 
is thus zero and  
$\sigma_{xy}$ for the model Eq.~\eqref{eq: effH} in the non-crossing approximation is equal to (see Fig.~\ref{fig: alldiagrams}, a)-c)):
\begin{align}
    \sigma^{\text{nc}}_{xy} =  - \frac{e^2}{h} \frac{m \epsilon_F }{\epsilon_F ^2 + m^2} \cdot n; \; \; \; n>1, \; |\epsilon_F|> |m|.
    \label{eq: nc}
\end{align}
Details about the evaluation of the diagrams
can be found in Appendix~\ref{ap: noncrosisotropic}.
Specifically, the side-jump contributes 
\begin{align}
    \sigma^{\text{sj}}_{xy} = -\frac{e^2}{h} \frac{m (\epsilon_F ^2 -m^2)}{2 \epsilon_F (\epsilon_F ^2 + m^2)}\cdot  n; \; \; \; n>1.
    \label{eq: sj}
\end{align}

\subsection{Diffractive Skew scattering}

We next consider diffractive
skew scattering which  originates from pairs of impurities separated by distances on the order of, or smaller than, the Fermi wavelength. In this regime, the scatterers are no longer amenable to a quasiclassical treatment as independent centers. As a result, a rigorous theoretical description of such correlated impurity scattering necessitates inclusion of both X and $\Psi$ diagrammatic contributions. (Fig.~\ref{fig: alldiagrams}, d)). However, the inclusion of three or more intersecting impurity lines in the conductivity diagrams yields an additional smallness, (maximally crossed diagrams 
contribute to a weak localization correction characterized by $\sigma_{xy} \sim \alpha$). 

Since the two impurities composing X or $\Psi$ are close by on the order of $\lambda_F$, we may evaluate internal lines with bare Green's functions, Eq.~\eqref{eq: momentumGF} at $\alpha = 0$. In real space this leads to 
\begin{align}
 G_0 ^{R/A} (\epsilon_F, \mathbf{r}) &= [\epsilon_F + \vec d(- i \nabla)]  \frac{\pi  \rho_n (\epsilon_F)}{2 \epsilon_F} \cdot \notag \\
& \left[\tilde Y_0 (p_0 r) \mp i  J_0 (p_0 r)\right]. 
\label{eq: cleanGF}
\end{align}
where
\begin{align}
    &\tilde Y_l (r) = Y_l (r) -\frac{2}{\pi} \sum_{j = 1}^{n - 1}  e^{-i \frac{ \pi j}{n}(2+l)} i^l K_l \left(  ie^{-i \frac{ \pi j}{n}} r\right),
    \label{eq: tildey} 
\end{align}
$J_l$ and $Y_l$ are the l-th Bessel functions of the first and
second kind, respectively; $K_l$ is the modified Bessel function of the second kind (cf.~Appendix~\ref{ap: GF_isotropic}).
We also used the density of states for isotropic model  $\rho_n(\epsilon_F)$:
\begin{align}
    &\rho_n(\epsilon_F) =\frac{\epsilon_F}{2\pi n v^{2/n}(\epsilon_F^2 - m^2)^{1 - 1/n}}.
    \label{eq: dos}
\end{align}

Using the real space representation, we obtain
\begin{subequations}
    \begin{align}
&\sigma_{xy}^{\text{X}} = 
 \frac{e^2}{h} \frac{ \epsilon_F m (\epsilon_F ^2 - m^2)}{(\epsilon_F ^2 +m^2)^2}   \pi n  \, \times \label{eq: xfull}  \\
&
\int_0^\infty dx \, J_1 J_n  [J_n \Tilde Y_0  - J_0  \Tilde Y_n], \notag \\
&\sigma_{xy}^{\Psi} = \frac{e^2}{h} \frac{ \epsilon_F m (\epsilon_F ^2 - m^2)}{(\epsilon_F ^2 +m^2)^2}  \pi n  \, \times \label{eq: psifull}  \\
&
\int_0^\infty dx \, J_1 J_n  [J_n \Tilde Y_0 + J_0  \Tilde Y_n] + J_0 J_n \Tilde Y_0 (J_{n-1}-J_{n+1}), \notag
\end{align}
\end{subequations}
 where for the
sake of convenience, we use the following compact notations $J_i = J_i (x), \, \Tilde Y_i = \Tilde Y_i (x)$, $x = p_0 R$, $R$ is a distance between two impurities.
For detailed calculations use Appendix~\ref{ap: dif_isotropic}. While 
Eqs.~\eqref{eq: xfull}-\eqref{eq: psifull} 
determine contributions from diffractive skew-scattering contribution for any $n\geq 1$, the analytical evaluation of these integrals can be obtained for $n = 1$~\cite{ado2015anomalous}, $n = 2$ (see Appendix~\ref{ap: bessels}), as well as in the limit $n \to \infty$: 
\begin{align}
    &\frac{\sigma_{xy}^{\text{X}}+ {\sigma_{xy}^{\Psi}}}{f(\epsilon_F, m)} \simeq  \frac{\sqrt{3}}{2\sqrt{2}} + \frac{4}{\pi^2} \begin{cases} 1 & n \text{ odd,} \\
    \frac{n^4}{(n^2-1)^2}, & n \text{ even} 
    \end{cases}.
    \label{eq: largen}
\end{align}

Numerical evaluations of the Bessel integrals are, of course, always possible and
for the most relevant experimental situations  
$n \in [2; 5]$ of 
$\sigma_{xy}^{\text{X}/\Psi} $  
are given in Table~\ref{tab:sigma_prefactors}. 

\begin{table}
\centering
\begin{tabular}{ c c c c c}
\toprule
 & \( n = 2 \) & \( n = 3 \) & \( n = 4 \) & \( n = 5 \) \\
\midrule
\( \sigma_{xy}^{\text{X}}/f(\epsilon_F, m) \)    & 0.719  & 0.563  & 0.542 & 0.530 \\
\( \sigma_{xy}^{\Psi}/f(\epsilon_F, m) \)        &0.466 & 0.509 & 0.494 & 0.506 \\
\bottomrule
\end{tabular}
\caption{\justifying 
 Prefactors of diffractive skew scattering contribution \( \sigma_{xy}^{\text{X}/\Psi} \) for different \( n \). Here, \( f(\epsilon_F, m)=\frac{e^2}{h}  \frac{\epsilon_F m (\epsilon_F^2 - m^2)}{ (\epsilon_F^2 + m^2)^2} \):  }
\label{tab:sigma_prefactors}
\end{table}

\subsection{Summary for weak dense impurities}

\begin{figure}
    \centering
\includegraphics[width=1\linewidth]{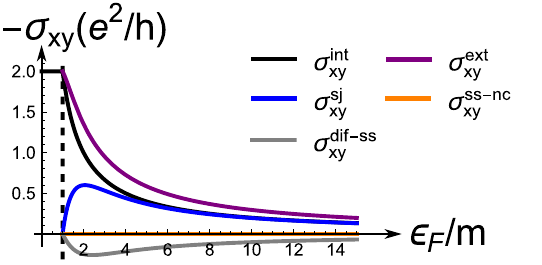}
    \caption{\justifying Intrinsic Eq.~\eqref{eq: int}, side-jump Eq.~\eqref{eq: sj}, skew-scattering
components Eqs.~\eqref{eq: xfull}-\eqref{eq: psifull}  and the extrinsic part of the anomalous Hall conductivity for $n=4$. We plot separately the skew scattering parts in the non-crossing approximation (purple) and diffractive (gray). The vertical dashed line marks \(\epsilon_F/m = 1\). All curves are shown for a single spin- and valley-polarized flavor.}
    \label{fig: components}
\end{figure}

 The full result for the anomalous Hall conductivity in the
zeroth order in $\alpha$ from one valley and one spin is given by Eqs.~\eqref{eq: nc}, \eqref{eq: xfull}-\eqref{eq: psifull}: 
\begin{align}
    \sigma_{xy}=\sigma_{xy}^{\text{nc}}+\sigma_{xy}^{\text{X}}+\sigma_{xy}^{\Psi}.
    \label{eq: fullisotropic}
\end{align}

When we compare these results with the well-known limiting case of $n = 1$, which corresponds to 2D massive Dirac
fermions \cite{ado2015anomalous, nagaosa2010anomalous,sinitsyn2007anomalous}, we recover the results 
of previous works if we artificially removed the vertex correction (which is non-vanishing for $n = 1$), see Eq.~\eqref{eq: correction_isotropic}.

The comparison of different components for anomalous Hall conductivity, Fig.~\ref{fig: components}, 
 for tetralayer graphene, illustrates that for Fermi energy sufficiently far inside the band the impurity scattering is comparable to the intrinsic contribution, yet upon lowering the Fermi energy,
the intrinsic mechanism dominates the anomalous Hall conductivity and ultimately is the sole contribution to the quantized plateau. Disorder broadens the band edge and may generate exponentially rare in-gap Lifshitz-tail states, which rounds the onset of the intrinsic contribution near the gap. Our results are controlled for $p_F l\gg 1$; the detailed band-edge physics for $p_F l\sim O(1)$ is nonperturbative and beyond our quantitative scope, while the plateau is expected to persist up to exponentially small corrections for weak disorder.
This is consistent with the
topological origin of the intrinsic contribution, which is
less sensitive to disorder. These results underscore that
in ABC-stacking graphene, the anomalous Hall effect is
largely determined by the interplay between intrinsic,
side-jump, and diffractive skew-scattering mechanisms,
while conventional skew-scattering may be suppressed.

\subsection{Strong sparse impurities}

 In the rest of the part about an isotropic model, 
 we consider the higher-moment 
 contribution of the disorder distribution 
  Eq.~\eqref{eq: nongaus} 
  to the first order in the parameter $\beta$ (third moment) given by 
  'Mercedes star' diagram(Fig.~\ref{fig: alldiagrams}. e)). 
We obtain a vanishing contribution for $n\geq 2$
\begin{align}
    \sigma_{xy}^{\text{MS}}&= \sigma_{xy}^{\rotatebox[origin=c]{360}{Y}} + \sigma_{xy}^{\rotatebox[origin=c]{180}{Y}} = \frac{e^2}{h} (\beta \upsilon^3 p_0 ^{3n-4})  \times \notag \\
    & \int_{\{\mathbf{p}_i\}} \Bigg (\text{Tr} [\mathbf{G}_{\mathbf{p_1}} ^A  j_x (\mathbf{p_1}) \mathbf{G}_{\mathbf{p_1}} ^R \mathbf{G}_{\mathbf{p_3}} ^R  \mathbf{G}_{\mathbf{p_2}} ^R  j_y (\mathbf{p_2}) \mathbf{G}_{\mathbf{p_2}} ^A 
 ] + \notag \\
 &\text{Tr} [\mathbf{G}_{\mathbf{p_1}} ^A  j_x (\mathbf{p_1}) \mathbf{G}_{\mathbf{p_1}} ^R \mathbf{G}_{\mathbf{p_2}} ^R  j_y (\mathbf{p_2}) \mathbf{G}_{\mathbf{p_2}} ^A 
 \mathbf{G}_{\mathbf{p_3}} ^A ]\Bigg) = 0,
 \label{eq: mercedes}
\end{align} 
because of angle integration. It should be noted that the higher moment gives us the same result 
(see Appendix~\ref{ap: thirdmoment}). At the same time  
note that a non-zero result does exist for $n=1$ \cite{sinitsyn2007anomalous}.

\section{Warped model}
\label{sec: warped}

This section addresses the 
effect of trigonal warping. 
In experimentally relevant regimes, the displacement field energy is comparable to the characteristic energy associated with the dispersion anisotropy, i.e.~warping. This implies that the isotropic calculation may be at the limit of its applicability, in particular at low doping,
 thus necessitating the  
 inclusion of warping in the analysis. 
 
 In the presence of warping in Eq.~\eqref{eq: effH}, i.e $w \neq 0$, the 
 system has $C_3$ symmetry instead of full-rotation symmetry as in isotropic model. It should be emphasized that this effective model is applicable for 
$n \geq 3$ and microscopically derives from 
various types of subleading interlayer hopping that were neglected in the isotropic model. 
 
As suggested by its name, the parameter $w$ introduces a warped Fermi surface. The 
approximate 
$\phi$ dependent Fermi momentum $p_F(\phi) \equiv p_\phi$ to second order in the 
power of warping parameter $\tilde w = w/v p_0^3$ is
\begin{equation}
    p_\phi = p_0 \left (1 -\frac{\tilde w \cos (3 \phi )}{n}+\frac{\tilde w^2 ((2 n-5) \cos (6 \phi )-5)}{4 n^2}\right ).
    \label{eq: fermimom1}
\end{equation}
We emphasize that at smallest doping $\tilde w$ is always large and our perturbative scheme breaks down. At the same time, it is worthwhile to  
note that Eq.~\eqref{eq: fermimom1}, provides a quantitatively accurate description of the Fermi surface (within few percents) for all $\tilde w < 1$ (Appendix \ref{sec: necdesignations}).

For the following calculations  
we define a $\phi$ dependent 
DOS
\begin{align}
    &\rho_{n,\phi}(\epsilon_F) = \epsilon_F \frac{p_\phi}{2n\pi f_\phi(p_\phi)} ,
\label{eq: DOSw}
\end{align}
where 
\begin{align}
&f_\phi(p_\phi) = \frac{dF_\phi(p)}{dp}; \; F_\phi(p) = \frac{\vert \vec d(p, \phi) \vert^2-\epsilon_F^2}{2n}.
\label{eq: Ff}
\end{align}

Our focus is on how 
 warping impacts the anomalous Hall conductivity, and we follow the same computational steps as in the isotropic model.
From now on, we will focus on the warping effect for trilayer and tetralayer graphene, i.e.~$n=3, 4$. However, similar reasoning can be used for other $n$.

\subsection{Intrinsic contribution}

We  
derive the intrinsic contribution for the warped model (cf.~Appendix~\ref{sec: intn3w},~\ref{sec: intn4w}): 
\begin{subequations}
\begin{align}
     &\sigma_{xy} ^{\text{w, int}} \vert^{n=3} =-\frac{e^2}{h}  \frac{m}{ \epsilon_F} \frac{3}{2 } \mathcal J_3^{(4)},
     \label{eq: warped_int_3}
\end{align}
\begin{align}
    &\sigma_{xy} ^{\text{w, int}} \vert^{n=4} =-\frac{e^2}{h}  \frac{m}{ \epsilon_F}  \cdot \left(2 \mathcal J_4^{(6)} +\tilde w \mathcal I_4^{(3)} + \frac{\tilde w^2}{8} \mathcal I_4^{(0)}  \right ).
    \label{eq: warped_int_4}
\end{align}
\end{subequations}
Here we introduced 
\begin{align}
    &  \mathcal I_n^{(l)} = \left \langle \bar p_\phi^l e^{i l \phi}\frac{\rho_{n,\phi}(\epsilon_F)}{\rho_n(\epsilon_F)} \right\rangle_\phi; \;
    \mathcal J_n^{(l)}  = \Big \langle\bar p_\phi^l \frac{\rho_{n,\phi}(\epsilon_F)}{\rho_n (\epsilon_F)}\Big \rangle_\phi; \;\notag \\
    &\bar p = p/p_0,
    \label{eq: integrals}
\end{align}
and $\langle \dots \rangle_\phi = \int_0^{2\pi} \dots d\phi$ denotes angular average.
These and the following integrals in the warped case can be evaluated numerically and will be discussed in the next section. The isotropic limit ($w=0$) can always be taken analytically and  
we can readily verify the agreement with our results for the isotropic results, in this case Eq.~\eqref{eq: int}.

\subsection{Self-energy}

We rely on the fact that the self-energy obtained in the self-consistent Born approximation is identical to that in the Born approximation
\begin{align}
    \Sigma^R 
    &\simeq \frac{\epsilon_F}{n \rho_n(\epsilon_F)} \alpha\int \frac{dp^2}{(2\pi)^2}\frac{\epsilon_F + \vec d(\v p) \cdot \sigma}{\epsilon_F^2 - \vec d(\v p)^2}.
\end{align}
For $n = 4$ and for $n$ not a multiple of 3, only the $\sigma_{0,z}$ components survive the angular integral due to $C_3$ symmetry.
For $n=3$ and for $n$ multiple of 3 we have all components. So, in general (see Appendix~\ref{sec: necdesignations}).
\begin{subequations}
\begin{equation}
        \text{Im} \Sigma^R  = - (\Gamma + \Gamma_x \sigma_x + \Gamma_y \sigma_y + \Gamma_z \sigma_z),
\end{equation}
where for $n=3,4$
\begin{align}
    &\Gamma  = -  \frac{\pi\alpha}{2n} \epsilon_F \mathcal I_n^{(0)}, \\
    &\Gamma_x  = - \delta_{n,3}\frac{\pi\alpha}{2n} v p_0 ^3 \cdot (\mathcal I_n^{(3)}+ \tilde w \mathcal J_n^{(0)}), \\
    &\Gamma_y  = 0,\\
    &\Gamma_z  = -  \frac{\pi\alpha}{2n} m \mathcal I_n^{(0)}. 
\end{align}
\label{eq: gammas}
\end{subequations}

\subsection{Non-crossing contribution: vertex corrections}

To compute the non-crossing and diffractive skew-scattering contributions, we need to include vertex corrections. Contrary to the isotropic model they generally do not vanish. We define $j_{\pm} = [{j_x \pm i j_y}]/{2}$, and use boldface symbols 
$\mathbf j_{\pm}$ for dressed vertices. 
We obtain (cf.~Appendix~\ref{sec: vert3w},~\ref{sec: vert4w}):
\begin{align}
    &\mathbf j_\pm \vert^{n=3} = j_\pm \vert^{n=3},   \\
    &\mathbf j_\pm \vert^{n=4} = j_\pm \vert^{n=4} \notag \\
    &+\left [4 vp_0^{3} \frac{\mathcal I_4^{(3)}}{\mathcal I_4^{(0)}} +  w \right ]  \frac{A}{1-A} \pm i \left [4 vp_0^{3} \frac{\mathcal I_4^{(3)}}{\mathcal I_4^{(0)}} +  w \right ]   \frac{ B}{(1- A)^2},
    \label{eq: correction_warped}
\end{align}
where
\begin{align}
    A = \frac{\epsilon_F^2 - m^2}{2(\epsilon_F^2 + m^2)}, \; B=- \frac{\epsilon_F m}{\epsilon_F^2 + m^2}  \frac{\pi \alpha}{n} \mathcal I_n^{(0)}.
\end{align}
So, the vertex correction in warped model for $n=3$ is zero 
but not for $n=4$.

Then, $\sigma_{xy}$ for warped model in the non-crossing approximation equals to (Appendix~\ref{sec: non3w},~\ref{sec: non4w}):
\begin{align}
    &\sigma_{xy}^{\text{w, nc}}\vert^{n=3} =- \frac{e^2}{h} \frac{3 m \epsilon_F \mathcal{I}_3^{(0)}}{\epsilon_F^2 - m^2} \int_0^{2 \pi} \frac{d \phi}{2 \pi} \frac{   \bar p_{\phi}^5}{ f_{\phi} (\bar p_{\phi})} \cdot \notag\\
    & \frac{1}{\mathcal{I}_3^{(0)} \frac{\epsilon_F^2 + m^2}{\epsilon_F^2 - m^2} + (\bar p_{\phi}^3 \cos{(3 \phi)+ \tilde w} )(\mathcal{I}_3^{(3)} + \tilde w \mathcal{J}_3^{(0)}) }, 
    \label{eq: warped_nc_3}
\end{align}
\begin{align}
    &\sigma_{xy}^{\text{w, nc}}\vert^{n=4} = - \frac{e^2}{h} \frac{4 m \epsilon_F}{\epsilon_F^2 + m^2}  \Big \{\left [{\frac{\mathcal J_4^{(6)}}{\mathcal I_4^{(0)}}} - \frac{[\mathcal I_4^{(3)}]^2}{[\mathcal I_4^{(0)}]^2} \right]   + \notag\\
    &\frac{4 \left(\epsilon_F ^2 + m^2\right)^2}{\left(\epsilon_F ^2 + 3 m^2\right)^2} \left [ \frac{\mathcal I_4^{(3)}}{\mathcal I_4^{(0)}}  + \tilde w/4\right ]^2\Big \}.
    \label{eq: warped_nc_4}
\end{align}
In the isotropic limit we recover Eq.~\eqref{eq: nc}.

\subsection{Diffractive Skew Scattering}

Finally, we can obtain diffractive skew-scattering contribution:
\begin{subequations}
\begin{align}
    \sigma_{xy}^{\text{w, X}}\vert^{n=3} =& - \frac{e^2}{h} \frac{m \epsilon_F }{\epsilon_F^2 - m^2} \cdot N_{X,3}, 
    \label{eq: wx3}\\
    \sigma_{xy}^{\text{w, X}}\vert^{n=4} =& -\frac{e^2}{h}  \frac{m \epsilon_F (\epsilon_F^2 -m^2)}{(\epsilon_F^2 + m^2)^2} \cdot N_{X,4}\vert^{1)} \notag \\
    &-\frac{e^2}{h}  \frac{m \epsilon_F (\epsilon_F^2 - m^2)^2}{(\epsilon_F^2 + 3 m^2)(\epsilon_F^2+m^2)^2} \cdot N_{X,4}\vert^{2)} 
    \notag\\
    &-\frac{e^2}{h}  \frac{m \epsilon_F (\epsilon_F^2 - m^2)^3}{(\epsilon_F^2 + 3 m^2)^2(\epsilon_F^2+m^2)^2} \cdot N_{X,4}\vert^{3)},
    \label{eq: wx4}
\end{align}
\begin{align}
    \sigma_{xy}^{\text{w}, \Psi}\vert^{n=3} =& -\frac{e^2}{h} \frac{m \epsilon_F }{\epsilon_F^2 - m^2} \cdot N_{\Psi,3}, 
    \label{eq: wpsi3}\\
    \sigma_{xy}^{\text{w}, \Psi}\vert^{n=4} =& -\frac{e^2}{h}  \frac{m \epsilon_F (\epsilon_F^2 -m^2)}{(\epsilon_F^2 + m^2)^2}  \cdot N_{\Psi,4}\vert^{1)} \notag \\
    &-\frac{e^2}{h}  \frac{m \epsilon_F (\epsilon_F^2 - m^2)^2}{(\epsilon_F^2 + 3 m^2)(\epsilon_F^2+m^2)^2} \cdot N_{\Psi,4}\vert^{2)} \notag\\
    &-\frac{e^2}{h}  \frac{m \epsilon_F (\epsilon_F^2 - m^2)^3}{(\epsilon_F^2 + 3 m^2)^2(\epsilon_F^2+m^2)^2}\cdot N_{\Psi,4}\vert^{3)}.
    \label{eq: wpsi4}
\end{align}
\label{eq: warped_dif}
\end{subequations}

The dimensionless integrals $N_{X,n}, N_{\Psi,n}$ over three angular variables, are too cumbersome to represented here, see Eqs.~\eqref{eq: NX3},~\eqref{eq: NPsi3},~\eqref{eq: NX41}-~\eqref{eq: NX43},~\eqref{eq: NPsi41}-~\eqref{eq: NPsi43} of the Appendix~\ref{ap: warped}. 

\subsection{Comparison isotropic and warped models}
\label{sec: comparison}

 In this section, we compare analytical results for the isotropic model with numerical results for the warped model and analyze the effect of trigonal warping on the anomalous Hall conductivity.

\begin{figure}
  \centering
  \begin{minipage}[t]{0.48\textwidth}
    \centering
    \includegraphics[width=\linewidth]{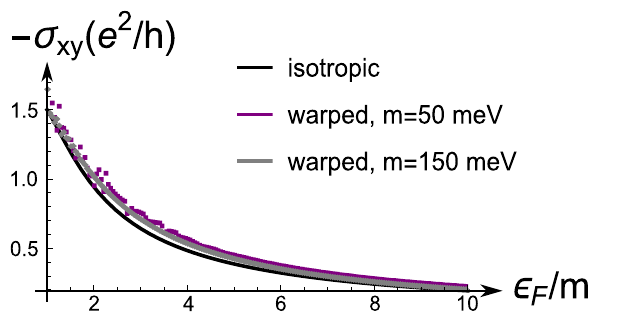}
    \vspace{0.3em}
    {\small (a)}
    \label{fig: fullN3}
  \end{minipage}\hfill
  \begin{minipage}[t]{0.48\textwidth}
    \centering
\includegraphics[width=\linewidth]{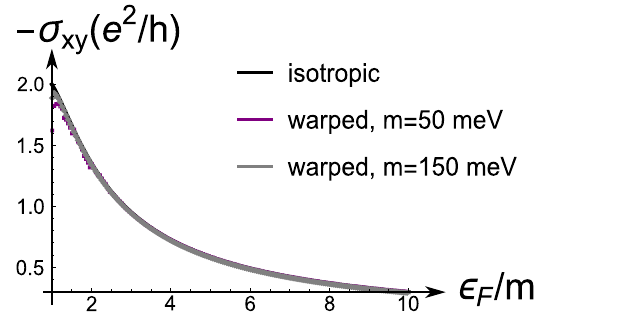}
    \vspace{0.3em}
    {\small (b)}
        \label{fig: fullN4}
  \end{minipage}
  \caption{\justifying Full anomalous Hall conductivity (one valley and one spin) for a) $n=3$
 with $\tilde w_3 = \frac{w}{m} \left(\frac{\epsilon_F ^2}{m^2}-1 \right)^{-1/2}$ and b) $n=4$ with warping parameter $\tilde w_4 = \frac{w}{\u^{1/4} m^{3/4}} \left(\frac{\epsilon_F ^2}{m^2}-1 \right)^{-3/8}$.}
  \label{fig: full}
\end{figure}

Importantly, in the warped model, all contributions depend on $\tilde w = \frac{w}{v p_0^3}$, it means that $\tilde w$ is a function of ($\epsilon_F, m$).
Thus, we numerically evaluate angular integrals and plot the full anomalous Hall conductivity as a function of $\epsilon_F/m$, at fixed displacement field $m$, comparing the isotropic and warped models for several values of the warping parameter: Fig.~\ref{fig: full}. There we used ab-initio estimates of the parameter $w/\u \approx 9.97 \times 10^{-3} \text{nm}^{-3}$ of the dispersion relation of graphene and 'warped' means the model includes the trigonal-warping term of amplitude $w = 20$ meV for $n = 3$ and $w = 34$ meV $\cdot$ nm for $n = 4$, while for 'isotropic' we set $w=0$ keeping all other parameters unchanged. 

These results suggest that the presence of warping increases the absolute value of the anomalous Hall conductivity in trilayer graphene, while in tetralayer graphene, it leads to only a slight decrease at low energies. In either case, within the present parameter regime, the impact of warping may be quantitatively relevant, but not qualitatively.

\section{Conclusion}
\label{sec: conclusion}
In conclusion, we have developed a theoretical framework to describe the anomalous Hall response in graphene samples with rhombohedral stacking. Within our approach we carefully include disorder scattering in a controlled fashion. In particular, we go beyond the conventional non-crossing approximation by incorporating crossed impurity-line diagrams, enabling a 
complete treatment of impurity scattering effects. We derived an analytical expression for an isotropic model that accounts for all key contributions to the anomalous Hall effect.
In addition, we investigated the role of 
trigonal warping—via detailed semi-numerical analysis. 

While warping introduces only a minor correction deep inside the band, impurity scattering has a pronounced impact on the behavior of 
$\sigma_{xy}$ only away from the band gap. 

Despite its simplicity, the isotropic model captures the essential physical features and might offer  
a practical and accurate description of rhombohedral graphene in the limit of sufficiently large displacement fields. The insights gained here may extend to a broader class of multiband materials or nonlinear extensions of the calculations done here \cite{du2019disorder,Kaplan2023scipost,kaplan2023general,kaplan2025quantumgeometricphotocurrentsquasiparticles} , enhancing our understanding of anomalous transport and guiding future experimental and theoretical studies. Our results also point to experimentally testable signatures.
While the intrinsic term yields a characteristic density dependence controlled by the Berry curvature of the underlying bands, extrinsic terms can introduce additional contributions that are sensitive to the disorder strength and microscopic disorder statistics. One may discriminate the two by varying regimes from ballistic to diffusive by altering sample-dimensions and or controlled addition of impurities.

\acknowledgements

It is a pleasure to thank Pavel Ostrovsky and Alex Levchenko for useful discussions. 
Support for this research was provided by the Office of the Vice Chancellor
for Research and Graduate Education at the University
of Wisconsin–Madison with funding from the Wisconsin
Alumni Research Foundation. This research was supported in part by grants NSF PHY-1748958 and PHY-2309135 to the Kavli Institute for Theoretical Physics
(KITP). DK, DG and EJK acknowledges hospitality
by the KITP, where this project was initiated. VM and EJK acknowledge hospitality by the Max-Planck-Institute for Solid State Research. 
DK is supported by the Abrahams Postdoctoral Fellowship of the Center for Materials Theory, Rutgers University and the Zuckerman STEM fellowship.
\bibliography{references}

\newpage

\appendix
\section{The low-energy Hamiltonian}
\label{ap: Hamiltonian}
In this appendix we provide details in the derivation of Eq.~\eqref{eq: effH} of the main text, i.e.~an effective
two-component Hamiltonian that describes hopping between
atomic sites $A_1$ and $B_n$.

We start from the Hamiltonian  in the basis $\Phi = [\phi_1, \phi_2, ..., \phi_n]^T$ with $\phi_l = [\psi_{A_l}, \psi_{B_l}]$ where $l$ is the layer number, $n$ is the number of layers, $\psi_A$ and $\psi_B$ are the weight on the $A, B$ sublattices:
\begin{equation}
\mathcal{H}_n =
\begin{pmatrix}
D_1   & V & W \\
V^\dagger   & D_2  & V & W \\
W^\dagger & V^\dagger   & D_3 & \ddots  \ddots \\
&  W^\dagger & \ddots & \ddots \ddots \\
&   & \ddots & \ddots & V & W \\
& & & V^\dagger & D_{n-1}  & V \\
& & & W^\dagger & V^\dagger & D_n 
\end{pmatrix},
\label{eq: H_N}
\end{equation}

\[
D_i  = \begin{pmatrix}
  U_i & \upsilon_F \mathcal{P}  \\
  \upsilon_F \overline{\mathcal{P}} & U_i 
 \end{pmatrix} ; \; V = \begin{pmatrix}
  -\upsilon_4 \mathcal{P} & \upsilon_3 \overline{\mathcal{P}}  \\
  \gamma_1 &   -\upsilon_4 \mathcal{P}
 \end{pmatrix}; \; W = \begin{pmatrix}
  0 & \gamma_2 /2  \\
  0 &  0
 \end{pmatrix},
\]
where $\upsilon_i = \sqrt{3} a \gamma_i /2 \hbar$ are effective velocities, $a = 0.246$ nm is the lattice spacing, $\upsilon_F = \sqrt{3} a \gamma_0 /2 \hbar$; $\mathcal{P} = \xi p_x - ip_y$ and $\xi = \pm 1$ is the valley index (we will keep $\xi =1$); $U_i$ describe on-site energies of the
atoms on layers in the presence of an external displacement field; $\gamma_0$ and
$\gamma_1$ describe nearest-neighbor intralayer and interlayer hopping respectively,
$\gamma_3$ represents hopping between
the low-energy sites of a $AB$-stacked bilayer, $\gamma_4$ couples low- and high-energy sites located on different layers, $\gamma_2$ describes the vertical coupling between sites $A_1$ and $B_3$ in a different unit cell. We will use the following values for energies: $\gamma_0 = 3.16$ eV, $\gamma_1 = 0.39$ eV, $\gamma_2 = -0.020$ eV, $\gamma_3 = 0.315$ eV, $\gamma_4 = 0.044$ eV \cite{koshino2009trigonal}.

We can derive an effective
two-component Hamiltonian for $n =3, 4$ that describes hopping between
atomic sites $A_1$ and $B_n$  using the procedure from Ref.~\cite{koshino2009trigonal}, where case for $n=3$ was considered:

\begin{widetext}
\begin{equation}
    \mathcal{H}_{2 \times 2 } (\mathcal{P})^{n=3} = \underbrace{\frac{\upsilon_F ^3}{\gamma_1 ^2}
\begin{pmatrix}
0 &\mathcal{P}^3 \\
\overline{\mathcal{P}}^3 & 0
\end{pmatrix}}_{H_3} + \underbrace{\left( \frac{\gamma_2}{2} - \frac{2 \upsilon_F \upsilon_3}{\gamma_1} p^2 \right) \sigma_x}_{H_w} + \underbrace{\frac{2 \upsilon_F \upsilon_4 p^2}{\gamma_1} \sigma_0}_{H_c} + \underbrace{\frac{U_1 - U_3}{2} \left(1 - \frac{\upsilon_F ^2 p^2}{\gamma_1 ^2} \right)\sigma_z}_{H_{a1}}  +
\label{eq: Hn3}
\end{equation}
\[\underbrace{\left(\frac{U_1 + U_3}{2} + \frac{2U_2 -U_1 -U_3}{2} \frac{\upsilon_F ^2 p^2}{\gamma_1 ^2} \right)\sigma_0}_{H_{a2}}, 
\]
\begin{equation}
    \mathcal{H}_{2 \times 2 } (\mathcal{P})^{n=4} =\underbrace{-\frac{\upsilon_F ^4}{\gamma_1 ^3}
\begin{pmatrix}
0 &\mathcal{P}^4 \\
\overline{\mathcal{P}}^4 & 0
\end{pmatrix}}_{H_4} + \underbrace{\left( \frac{3 \upsilon_3 \upsilon_F^2 }{\gamma_1^2} p^2 - \frac{\gamma_2 \upsilon_F }{\gamma_1}\right) \begin{pmatrix}
0 &\mathcal{P} \\
\overline{\mathcal{P}} & 0
\end{pmatrix}}_{H_w} + \underbrace{\frac{2 \upsilon_F \upsilon_4 p^2}{\gamma_1} \sigma_0}_{H_c} +
\label{eq: Hn4}
\end{equation}
\[\underbrace{\Bigg[\frac{U_1 - U_4}{2} \left(1 - \frac{\upsilon_F ^2 p^2}{\gamma_1 ^2}\right) + \frac{U_2 - U_3}{2} \frac{\upsilon_F^2 p^2}{\gamma_1^2} \left(1 - \frac{2\upsilon_F ^2 p^2}{\gamma_1 ^2}\right)\Bigg ]\sigma_z}_{H_{a1}}  + \underbrace{\Bigg[\frac{U_1 + U_4}{2} \left(1 - \frac{\upsilon_F ^2 p^2}{\gamma_1 ^2}\right) + \frac{U_2 + U_3}{2} \frac{\upsilon_F^2 p^2}{\gamma_1^2} \Bigg ]\sigma_0}_{H_{a2}},\]
\end{widetext}

where $H_n$ describes hopping between sites $A_1$ and $B_n$, $H_w$ is the term which correspond to warping, $H_c$ introduces electron-hole asymmetry, $H_{a1}$ and $H_{a2}$ stems from  the interlayer asymmetries. To simplify the effective Hamiltonian, which can be written as
\begin{equation}
    H = d_0(\v p) + \vec d(\v p) \cdot \vec \sigma,
\end{equation}
we consider the relative magnitude of the various terms in eqs.~\eqref{eq: Hn3}-\eqref{eq: Hn4}. 

In the mass term
\begin{equation}
    d_z(\v p) = m \left ( 1 + m_2 p^2 + (m_2 p^2)^2 + \dots. \right),
\end{equation}
terms involving $m_2 \sim v_F^2/\gamma_1^2 \sim 100 a^2$ are negligible so long as $n \lesssim \frac{1}{100 a^2} \sim 16 \times 10^{12} cm^{-2}$. In the intersublattice terms
\begin{equation}
    d_{x-iy}(\v p) = v \mathcal P ^n + w(\v p) \mathcal P^{n - 3},
\end{equation}
the term $w(\v p) = w(1 + w_2 p^2)$ contains $w_2 \sim \frac{v_F v_3}{\gamma_1 \gamma_2}$ is about a factor of 2 larger than the $m_2$ term. As such, it can be neglected so long as $n \lesssim \frac{1}{100 a^2} \sim 8 \times 10^{12} cm^{-2}$.

Finally, we compare $d_0(\v p) = \frac{v_F v_4 \v p^2}{\gamma_1}$ (the constant part is absorbed into the chemical potential) to  $\vert \vec d(\v p) \vert \sim m$ so that 
\begin{equation}
    \frac{d_0}{\vert \vec d \vert} =  \underbrace{\frac{v_F^2 p^2}{\gamma_1^2} }_{ = m_2 p^2 }\underbrace{\frac{v_4 }{v_F}}_{=1/100} \underbrace{\frac{\gamma_1}{m}}_{ = 4 eV/10 m}.
\end{equation}
We can estimate $m \sim eV$ so that $ \frac{d_0}{\vert \vec d \vert} \sim m_2 p^2 \frac{1}{250}$ so that $d_0(\v p)$ can be dropped wherever we dropped $m_2$. 
Typical densities in experiments are a fraction  of $10^{12}cm^{-2}$. 
This concludes the derivation of Eq.~\eqref{eq: effH}
\subsection{Importance of impurities}
\label{sec: impur}
In this section, we discuss when impurities are  important in experimental samples and comment on the values of $\alpha$ and $\beta$.

First, the experimental article \cite{zhou2021superconductivity} quotes a mean free path $ l\sim 1$ $\mu$m (it was deduced from a Drude expression for $R_{xx}$) at density $n_e \approx 2 \cdot 10^{12}$ cm$^{-2}$ and the sample size $L\sim 2.5$ $\mu$m. 
Alternatively, as quantum oscillations \cite{zhou2021half} become visible
for a magnetic field $\sim.1$ T, we estimate a mean-free path $l \sim 100$ nm at density $n_e \approx 0.25 \cdot 10^{12}$ cm$^{-2}$. 

To connect these experimental numbers to our disorder parametrization, we recall that $\alpha$ and $\beta$ control the second and third moments of the random potential, respectively, Eqs.~\eqref{eq: gaus}--\eqref{eq: nongaus}. 
The parameter $\alpha$ sets the elastic scattering rate within SCBA. 
Using the standard Born estimate $1/\tau \sim 2\pi\,\nu(\epsilon_F)\langle V V\rangle$ (up to $O(1)$ factors), where $\nu(\epsilon_F)$ is the density of states at the Fermi energy, and $l=v_F\tau$, we obtain parametrically $\alpha\sim (p_0 l)^{-1}$. 
For the experimental mean free paths quoted above this yields $\alpha\sim 10^{-2}$--$10^{-3}$, consistent with the good assumption used in our analysis. 
In contrast, $\beta$ parametrizes the third cumulant (skewness) of the disorder distribution and cannot be extracted from $l$ alone; it vanishes for Gaussian disorder and is expected to be strongly suppressed for weak dense disorder, while it can become relevant in the strong-sparse, non-Gaussian regime. In simple models of weak dense disorder one expects $\beta$ to be parametrically smaller than $\alpha$ (e.g., $\beta \propto \alpha^{3/2}$ times a dimensionless skewness).

\section{Isotropic model}
\subsection{Real space clean Green's function}
\label{ap: GF_isotropic}
In this appendix we present details in the derivation of Eq.~\eqref{eq: cleanGF}. We adopt the convention $\epsilon \equiv \epsilon_F >0$ throughout the appendices to streamline the derivations.
We start from the definition of Green's function:
\begin{gather}
    G_0 ^{R/A} (\epsilon, \mathbf{p}) = [\epsilon \pm i0 - H_0]^{-1}, 
\end{gather}
\[
     G_0 ^{R/A} (\epsilon, \mathbf{r}) = 
[ \epsilon + \vec d(- i \nabla) ]  I_n (\mathbf{r}),
\]
where
\[ I_n (\mathbf{r}) =  \frac{ 2\pi n\rho_n(\epsilon)}{\epsilon }\underbrace{\int \frac{d \bar p_x d \bar p_y}{(2\pi)^2} \frac{e^{i \bar p_x p_0 r}}{1 \pm i 0 - (\bar p_x^2 + \bar p_y^2)^n}}_{\bar I_n(p_0 r)},
\]
where $p^2 = \bar p_x^2 + \bar p_y^2$.
We use:
\begin{equation}
    \frac{1}{1+i 0 -  p^{2n}} = \frac{1}{n} \sum_{j = 0}^{n - 1} \frac{1}{1+i0 - \zeta^j p^2}; \; \zeta = e^{i 2\pi/n},
    \label{eq: decGF}
\end{equation}
to define
\begin{align}
    &\bar I_n(p_0r) = \frac{1}{4n}\Bigg[Y_0 (p_0 r) \mp i J_0 (p_0 r) - \notag\\
    &-\frac{2}{\pi} \sum_{j = 1}^{n - 1}  e^{-i \frac{2 \pi j}{n}} K_0 \left(  ie^{-i \frac{ \pi j}{n}} p_0 r\right)\Bigg]. 
\end{align}

\subsection{Self-consistent Born approximation}
\label{ap: SCBA}

In this appendix we provide details in the derivation of Eq.~\eqref{eq: momentumGF}. The self-consistent Born approximation leads to a self-energy:
\begin{align}
    &\Sigma = \frac{\epsilon }{n \rho_n(\epsilon)} \alpha \cdot \notag\\
    &\int \frac{d^2 p}{(2\pi)^2} \frac{\epsilon + m \sigma_z + \upsilon \sigma_x p^n \cos(n \theta) + \upsilon \sigma_y p^n \sin(n \theta) }{\upsilon^2 p_0 ^{2n} - \upsilon^2 p^{2n} \pm i \Gamma (\epsilon + m)} = \notag\\
    &\frac{-\pi i \alpha (\epsilon + m\sigma_z) }{2n}   = -i (\Gamma + \Gamma_z \sigma_z)
\end{align}

Here we only kept the imaginary part of the self-energy, the real part is absorbed into the chemical potential and mass. From this, we can readily get Eq.~\eqref{eq: momentumGF}.

\subsection{Intrinsic contribution for isotropic model}
\label{ap: int_isotropic}
In this appendix, we show details for the intrinsic contribution, Eq.~\eqref{eq: int}, which in the Kubo-Streda approach is given by
\begin{gather}
\sigma_{xy} ^{\text{int}} = \frac{e^2}{h} \int \frac{dp^2}{(2\pi)^2} \text{Tr} \Bigg ( j_x (p) G_p ^{R}  j_y (p) G_p ^{A} \Bigg).
\end{gather}
After taking the trace we can get
\begin{widetext}
\begin{align*}
    \sigma_{xy} ^{\text{int}} = \frac{e^2}{h} \int_0^\infty \frac{pdp}{2\pi} \int_0^{2\pi} \frac{d\theta}{2\pi} \frac{\upsilon^2 n^2 (4\epsilon m i (i0) p^{2n-2} + i\upsilon^2 p^{4n-2} (e^{-2i\theta} - e^{2i\theta})) }{(\upsilon^2 p_0 ^{2n} - \upsilon^2 p^{2n})^2 - 4(i0)^2 \epsilon^4} = - \frac{e^2}{h} \frac{m }{2\epsilon} n.
\end{align*}
\end{widetext}

Consistently, using the Berry curvature $\Omega^{(n)}_{\pm}(\bm{p})=\pm \frac{1}{2}\frac{n^2 \upsilon^2 m p^{2n-2}}{(m^2 + \upsilon^2 p^{2n})^{3/2}} $, we obtain
\begin{align}
&\sigma_{xy} = \frac{e^2}{\hbar} \displaystyle\sum_{\pm} \int \frac{d^2 p}{(2\pi)^2} f_{FD} (\pm \xi - \epsilon) \Omega^{(n)}_{\pm} (\mathbf{p}) =
\notag \\
&-\frac{e^2}{\hbar}\frac{n}{2} \Big(\Theta(m-\epsilon) + \Theta(\epsilon-m)\cdot \frac{|m|}{\epsilon}\Big) \cdot \text{sgn}(m).
\end{align}
Here $\xi = |\vec d(\v p)|$.

\subsection{Vertex correction}
\label{ap: vertex}
In this appendix, we calculate a single-impurity vertex correction, which is given by Eq.~\eqref{eq: correction_isotropic} of the main text. Let us consider the first order correction:

\begin{align}
    &\delta j_{x/y} = \alpha \frac{\epsilon}{n \rho_n (\epsilon)} \int \frac{d^2 p}{(2\pi)^2} \mathbf{G}_{\mathbf{p}} ^A  j_{x/y}  \mathbf{G}_{\mathbf{p}} ^R \propto \notag\\
    & \int \frac{d^2 p}{(2\pi)^2} (a(|\vec {\mathbf{p}}|^2) + b(|\vec {\mathbf{p}}|^2) \P^n + c(|\vec {\mathbf{p}}|^2)\bar \P^n) \P^{n-1} + \text{h.c.}=0. 
    \label{eq: vc}
\end{align}
 The matrix-valued isotropic functions $a(|\vec {\mathbf{p}}|^2), b(|\vec {\mathbf{p}}|^2), c(|\vec {\mathbf{p}}|^2)$ may be read off from $\mathbf{G}_{\mathbf{p}} ^R$ given in Eq.~\eqref{eq: momentumGF}. 

Due to angle integration, all ladder diagrams will be equal to zero if $n$ is not equal to 1, as a consequence, the contribution from skew-scattering diagrams in non-crossing approximation is equal to zero too.

\subsection{$\sigma_{xy}$ in non-crossing approximation}
\label{ap: noncrosisotropic}
Now, we show details for the contribution from non-crossing diagrams of Eq.~\eqref{eq: nc}, which is  given by
\begin{widetext}
    \begin{align}
&\sigma^{\text{nc}}_{xy} = \frac{e^2}{h} \int \frac{d^2p}{(2\pi)^2} 
\text{Tr} \left( j_x(p) \mathbf{ G^R}(p) j_y(p) \mathbf{ G^A}(p) \right) =\notag \\
& \frac{e^2}{h} \int_0^\infty \frac{p dp}{2\pi} \int_0^{2\pi} \frac{d\theta}{2\pi} 
v^2 n^2 
\frac{-4\epsilon m \frac{\pi\alpha}{n} p^{2n-2} + i v^2 p^{4n-2} (e^{-2i\theta} - e^{2i\theta})}
{(v^2 p_0^{2n} - v^2 p^{2n})^2 + \frac{\pi^2 \alpha^2 (\epsilon^2 + m^2)^2}{n^2}}=- \frac{e^2}{h} \frac{m \epsilon n}{\epsilon^2 + m^2}.
\end{align}
\end{widetext}

The non-crossing contribution includes, in principle, the intrinsic, side-jump and Gaussian skew-scattering components. The latter vanishes, Eq.~\eqref{eq: correction_isotropic}, so we can easily find side-jump components:
\begin{align}
\sigma^{\text{sj}}_{xy} = \sigma^{\text{nc}}_{xy} - \sigma^{\text{int}}_{xy}= -\frac{e^2}{h} \frac{m (\epsilon^2 -m^2)}{2 \epsilon (\epsilon^2 + m^2)}\cdot  n; \; \; \; n>1.
\end{align}

\subsection{Diffractive skew-scattering}
\label{ap: dif_isotropic}
In this appendix we provide calculations about diffractive skew-scattering contribution from X and $\Psi$ diagram Eq.~\eqref{eq: xfull}, ~\eqref{eq: psifull} of the main text. 
\begin{widetext}
    \begin{align}
    \sigma_{xy}^\text{X} = \frac{e^2}{h} \Bigg(\frac{\epsilon }{n \rho_n(\epsilon)} \alpha \Bigg)^2 \int_{\{\mathbf{p}_i\}} \text{Tr} [\mathbf{G}_{\mathbf{p_1}} ^A  j_x (\mathbf{p_1}) \mathbf{G}_{\mathbf{p_1}} ^R \mathbf{G}_{\mathbf{p_3}} ^R \mathbf{G}_{\mathbf{p_2}} ^R j_y (\mathbf{p_2}) \mathbf{G}_{\mathbf{p_2}} ^A \mathbf{G}_{\mathbf{p_4}} ^A ] (2\pi)^2 \delta(\mathbf{p_1}+\mathbf{p_2}-\mathbf{p_3}-\mathbf{p_4}).
\end{align}
\end{widetext}

We rewrite the average Green's function  as 
\begin{align}
    &\mathbf{G}_{\mathbf{p}}^R = N(\mathbf{p}) \mathcal{G}_{\mathbf{p}}^R, \notag\\
&N(\mathbf{p}) = \epsilon + \vec{d}(\vec{\mathbf{p}})\vec{\sigma}, \; \mathcal{G}_{\mathbf{p}}^R = 
\frac{1}{\epsilon^2 - \xi^2 + \frac{i \pi \alpha }{n } (\epsilon^2 +m^2)}. \notag
\end{align}

In this notation, we use that within our approximation we can keep only the lowest order in $\alpha$ in $N(\mathbf{p})$, i.e.~zeroth order.
Then, using partial fraction decomposition and keeping overall $O(\alpha^0)$
\begin{widetext}
    \[\sigma_{xy}^{\text{X}} = \frac{e^2}{h} \Bigg(\frac{\epsilon }{n \rho_n(\epsilon)} \alpha \Bigg)^2  \cdot {\frac
{n^2}{\pi^2 \alpha^2 (\epsilon^2 +m^2)^2}}\cdot  \int_{\{\mathbf{p}_i\}}  (2\pi)^2 \delta(\mathbf{p_1}+\mathbf{p_2}-\mathbf{p_3}-\mathbf{p_4}) \cdot
\]
\[
\Bigg(\text{Tr}  [N(\mathbf{p_1})  j_x (\mathbf{p_1}) N(\mathbf{p_1}) N(\mathbf{p_3})  N(\mathbf{p_2}) j_y (\mathbf{p_2}) N(\mathbf{p_2})  N(\mathbf{p_4})]\cdot \mathfrak{I}\mathcal{G}_{\mathbf{p_1}}\mathfrak{I}\mathcal{G}_{\mathbf{p_2}} (\mathfrak{R}\mathcal{G}_{\mathbf{p_3}} - i\mathfrak{I}\mathcal{G}_{\mathbf{p_3}})(\mathfrak{R}\mathcal{G}_{\mathbf{p_4}} + i\mathfrak{I}\mathcal{G}_{\mathbf{p_4}}) \Bigg),
\]
\end{widetext}
where $\mathfrak{R}$ and $\mathfrak{I}$ denote real and imaginary parts. 

After taking the trace and nondimensionalization, we can rewrite it using Green's functions in real space, App.~\ref{ap: GF_isotropic}: 
\begin{align*}
    &\sigma_{xy}^{\text{X}}  = \Xi_n \int_{\ \{\mathbf{r}_i\}, \{\mathbf{R}\}} T_{\text{X}} \cdot \mathfrak{I}(r_1)\mathfrak{I}(r_2)\mathfrak{I}(r_3)\mathfrak{R}(r_4)\cdot\\
    &\delta(\mathbf{R-r_1}) \delta(\mathbf{R-r_2}) \delta(\mathbf{R+r_3}) \delta(\mathbf{R+r_4}),
\end{align*}

where
\begin{align}
   & \Xi_n = -\frac{e^2}{h} \frac{m \epsilon (\epsilon^2 - m^2)}{8 (\epsilon^2 + m^2)^2}, \notag\\ 
   & T_{\text{X}} =
    \hat \P_1 ^{n-1} \hat {\bar{\P}}_2 ^{n-1}(\hat \P_2 ^n \hat {\bar{\P}}_1 ^n -1) [\hat {\bar{\P}}_1 ^n (\P_3 ^n - \P_4 ^n) + \P_2 ^n (\hat {\bar{\P}}_3 ^n -\hat {\bar{\P}}_4 ^n)] \notag
\end{align}

We use complex polar space coordinates $\hat \P = \hat p_x - i \hat p_y = - i e^{-i \phi} (\partial_r - \frac{i}{r}\partial_{\phi})$, so that 
\begin{align*}
    &\hat {\P}^m =(-i)^m e^{-i m \phi} \, \times \, \notag\\
    &\Bigg[\underbrace{\left(\partial_r - \frac{n-1}{r} - \frac{i}{r} \partial_{\phi}\right)... \left( \partial_r - \frac{1}{r} - \frac{i}{r} \partial_{\phi}\right) \left( \partial_r - \frac{i}{r} \partial_{\phi}\right)}_{m \text { differ. operators }}\Bigg]
\end{align*}

Finally, general answer for any $n$:
   \begin{align*}
&\sigma_{xy}^{\text{X}} = 
 \frac{e^2}{h} \frac{ \epsilon_F m (\epsilon_F ^2 - m^2)}{(\epsilon_F ^2 +m^2)^2} \cdot  \pi n  \, \times \notag \\
&
\int_0^\infty dx \, J_1 J_n  [J_n \Tilde Y_0  - J_0  \Tilde Y_n].
\end{align*}

 The contribution from the $\Psi$ diagram can be calculated in the same way as for the X diagram, so we can get the following:
\begin{align*}
    &\sigma_{xy}^{\Psi} = \frac{e^2}{h} \frac{ \epsilon_F m (\epsilon_F ^2 - m^2)}{(\epsilon_F ^2 +m^2)^2} \cdot  \pi n  \, \times  \\
&
\int_0^\infty dx \, J_1 J_n  [J_n \Tilde Y_0 + J_0  \Tilde Y_n] + J_0 J_n \Tilde Y_0 (J_{n-1}-J_{n+1}), \notag
\end{align*}

This concludes the derivation of Eqs.~\eqref{eq: xfull}-\eqref{eq: psifull} of the main text.
\subsection{Calculation of integral over Bessel functions}
\label{ap: bessels}
In this section, we present calculations of the Bessel-function integrals appearing in~\(\eqref{eq: xfull}\)–\(\eqref{eq: psifull}\) and obtain asymptotic expressions in the limit $n \to \infty$.

We need the following integrals:
\begin{subequations}
    \begin{align}
        &I_{n10n} = \int_0^\infty dx J_n J_1 J_0 \tilde Y_n, \\
        &I_{nn10}= \int_0^\infty dx J_n^2 J_1 \tilde Y_0, \\
        &I_{nn01} = \int_0^\infty dx J_n^2 J_0 \tilde Y_1, \\
        &I_{n \pm  00} = \int_0^\infty dx J_n (J_{n-1} - J_{n+1}) J_0 \tilde Y_0.
    \end{align}
    \label{eq: integrals}
\end{subequations}
The integral $I_{n10n}$ drops out of the final expression for the diffractive skew-scattering contribution and $I_{n \pm 00}$ can be expressed in terms of the others as $I_{n \pm00} = I_{nn10} +I_{nn01}$. Since $\tilde{Y}$ (see Eq.~\eqref{eq: tildey}) decomposes into a $Y$ term and a $K$ term, we split each integral accordingly, for example:
\begin{align}
    I_{nn10} = I_{nn10}^Y + I_{nn10}^K,
    \label{eq: sum}
\end{align}
where
\begin{align*}
    &I_{nn10}^Y = \int_0^\infty dx J_n^2 J_1 Y_0,\notag \\
    &I_{nn10}^K = \int_0^\infty dx J_n^2 J_1 {\left(\frac{2}{\pi} \sum_{j=1}^{n-1} (-e^{i \frac{2\pi j}{n}}) K_0 (i e^{-i \frac{\pi j}{n}} x)\right)}.
\end{align*}

The general method we use to handle these integrals is as follows \cite{sm}:
\begin{itemize}
    \item Reduce the number of Bessel functions from 4 to 3 using the following expression:
    \begin{align*}
    J_n^2 &=  \int_0^2\frac{  da \; x^n}{\pi^{1/2}\Gamma(n + 1/2)} \left (\frac{a}{2} \right)^n \left [1-\left (\frac
    {a}{2} \right)^2 \right]^{n-1/2} J_n(a x).
\end{align*}
\item Integrate over $x$ using one of these four identities ($u(a,c) = \frac{a^2 + c^2 - 1}{2a c}$): 
\begin{align*}
    I_{110}^Y(a,c) & = \int_0^\infty dx x J_1(a x) J_1(x) Y_0(c x) \\
    & = \frac{1}{\pi a } \begin{cases}
        1- \frac{u- a/c}{\sqrt{u^2 - 1}}, & 1< c - a, \\
        1+ \frac{u- a/c}{\sqrt{u^2 - 1}}, & 1< (a - c) \lor 1 > (a + c), \\
        1, & \vert a- c \vert <1< a + c,
    \end{cases}\\
      I_{101}^Y(a,c) & = \int_0^\infty dx x J_1(a x) J_0(x) Y_1(c x) \\
    & = - \frac{1}{\pi a c} \begin{cases}
        1- \frac{u}{\sqrt{u^2 - 1}}, & 1< c - a, \\
        1+ \frac{u}{\sqrt{u^2 - 1}}, & 1< (a - c) \lor 1 > (a + c), \\
        1, & \vert a- c \vert <1< a + c,
    \end{cases}\\
    I_{110}^K(a,c) & = \int_0^\infty dx x J_1(a x) J_1(x) K_0(c x) \\
    & = \frac{a^2+c^2+1}{2 a \sqrt{\left(a^2+c^2+1\right)^2-4 a^2}}-\frac{1}{2 a}, \\
     I_{101}^K(a,c) & = \int_0^\infty dx x J_1(a x) J_0(x) K_1(c x) \\
    & = \frac{a^2-c^2-1}{2 a c\sqrt{\left(a^2+c^2+1\right)^2-4 a^2}}+\frac{1}{2 a c}.
\end{align*}
\item Integrate over a. 
\end{itemize}

Employing this approach, we find 
\begin{align}
    &I_{nn10}^Y = \frac{1}{2 \pi n}; \: I_{nn01}^Y = -\frac{1}{2 \pi n}.
    \label{eq: I^Y}
\end{align}

Although obtaining $I^{K}$ in general is analytically prohibitive, the method outlined above allows us to derive the $n=2$ case:
\begin{widetext}
    \begin{align}
I_{2210}^K =& \frac{1}{4 \pi} \Bigg(  -\, _3F_2\left(\frac{1}{2},\frac{5}{4},\frac{7}{4};2,\frac{5}{2};-4\right)+ 2 \, _3F_2\left(\frac{1}{4},\frac{1}{2},\frac{3}{4};1,\frac{3}{2};-4\right)-1\Bigg), \\
    I_{2201}^K = &\frac{1}{4 \pi} \Bigg( 1 -2 \, _3F_2\left(\frac{1}{4},\frac{1}{2},\frac{3}{4};1,\frac{3}{2};-4\right) -_3F_2\left(\frac{1}{2},\frac{5}{4},\frac{7}{4};2,\frac{5}{2};-4 \right)+ 2 \, _3F_2\left(\frac{1}{2},\frac{3}{4},\frac{5}{4};\frac{3}{2}, 2;-4\right)\Bigg).
\end{align}

\end{widetext}

Thus, the $n=2$ diffractive skew-scattering contribution—accurate up to an overall prefactor—reads:
\begin{align}
    &\frac{\sigma_{xy}^{\text{X}}+ {\sigma_{xy}^{\Psi}}}{f(\epsilon_F, m)} =  3 \pi n I_{2210} + \pi n I_{2201} \approx 1.18617. \notag
\end{align}
These values agree with the values obtained by straightforward numerical calculation in Table~\ref{tab:sigma_prefactors} in the main text.  
 We next focus on $n\geq 3$ and develop a controlled large-$n$ expansion, yielding the $n\to\infty$ asymptotics.

We start from the integral \(I^{K}_{nn10}\).
After completing the first two steps of our approach, we obtain:
\begin{align}
    I_{nn10}^K &  = \int_0^2\frac{  da}{\pi^{1/2}\Gamma(n + 1/2)} \left (\frac{a}{2} \right)^n \left [1-\left (\frac
    {a}{2} \right)^2 \right]^{n-1/2} \notag \\
    & (-1)^{n-1}(\partial_a - \frac{n-1}{a})(\partial_a - \frac{n-2}{a}) \dots (\partial_a - \frac{1}{a}) \times \notag \\
    & \underbrace{{\frac{2}{\pi} \sum_{j = 1}^{n- 1} (-e^{- i \frac{2\pi j}{n}}) I_{110}^K(a, i e^{- i \frac{\pi j}{n}})}}_{\Sigma^{K_0}(a)}. 
\end{align}

First, we replace the discrete sum by a continuum integral
\begin{align}\label{eq:ApproxLargen}
    &\frac{1}{n} \sum_{j = 1}^{n} e^{- i \frac{2\pi j}{n}} I_{110}^K(a, i e^{- i \frac{\pi j}{n}} ) \rightarrow \int \frac{d\phi}{2\pi}e^{- i \phi}  I_{110}^K(a, i e^{- i \phi /2})  \notag \\
    & +\delta_n(a)\simeq -\frac{1}{\pi} \sqrt{1 - (a/2)^2} -\frac{\kappa}{\sqrt{a n}} e^{-2an},
\end{align}
where $\kappa$ is a dimensionless prefactor determined by comparison with numerical data ($\kappa \approx \sqrt{\pi}$); below we keep the $\sqrt{\pi}$ factor explicit.

Thus, for large $n$ we find:
\begin{align}
    \Sigma^{K_0}(a) \simeq -\frac{1}{\pi a} + \frac{2n}{\pi^2}\sqrt{1 - (a/2)^2} + \frac{2 \sqrt{n}}{\sqrt{a \pi}} e^{-2an}.
\end{align}

Then, after integrating over $a$, we obtain the $n\to\infty$ limit:
\begin{align}
    I_{nn10}^K \simeq -\frac{1}{2 \pi n} + \frac{2}{\pi^3} \begin{cases} \frac{1}{n}, & n \text{ odd,} \\
    \frac{n}{n^2-1}, & n \text{ even} 
    \end{cases} + \frac{\sqrt{3}}{4 \sqrt{2} \pi} \frac{1}{n}.
    \label{eq: I^K _0}
\end{align}
An analogous calculation for $I^{K}_{nn01}$ yields the $n\to\infty$ limit:
\begin{align}
    I_{nn01}^K \simeq \frac{1}{2 \pi n} - \frac{2}{\pi^3} \begin{cases} \frac{1}{n}, & n \text{ odd,} \\
    \frac{n(n^2-3)}{(n^2-1)^2}, & n \text{ even} 
    \end{cases} - \frac{\sqrt{3}}{4 \sqrt{2} \pi} \frac{1}{n}.
    \label{eq: I^K _1}
\end{align}

Combining Eqs.~\eqref{eq: integrals}–\eqref{eq: I^Y} and \eqref{eq: I^K _0}–\eqref{eq: I^K _1}, 
we obtain the diffractive skew-scattering contribution in the large-\(n\) limit, 
up to an overall prefactor:
\begin{align}
    &\frac{\sigma_{xy}^{\text{X}}+ {\sigma_{xy}^{\Psi}}}{f(\epsilon_F, m)} \simeq  \frac{\sqrt{3}}{2\sqrt{2}} + \frac{4}{\pi^2} \begin{cases} 1 & n \text{ odd,} \\
    \frac{n^4}{(n^2-1)^2}, & n \text{ even} 
    \end{cases}.
\end{align}

\subsection{Case of sparse and strong impurities}
\label{ap: thirdmoment}
 In this appendix we deduce the contribution of the higher-order moment, Eq.~\eqref{eq: nongaus}, of the disorder potential Eq.~\eqref{eq: mercedes}.

The contribution from "Mercedes star" diagram: 
\begin{align}
&\sigma_{xy}^{\text{MC}}=\sigma_{xy}^{\rotatebox[origin=c]{360}{Y}} + \sigma_{xy}^{\rotatebox[origin=c]{180}{Y}} = \frac{e^2}{h} (\beta \upsilon^3 p_0 ^{3n-4})  \cdot \notag\\
&\int_{\{\mathbf{p}_i\}} \Bigg (\text{Tr} [\mathbf{G}_{\mathbf{p_1}} ^A  j_x (\mathbf{p_1}) \mathbf{G}_{\mathbf{p_1}} ^R \mathbf{G}_{\mathbf{p_3}} ^R  \mathbf{G}_{\mathbf{p_2}} ^R  j_y (\mathbf{p_2}) \mathbf{G}_{\mathbf{p_2}} ^A 
 ] +\notag \\
 &\text{Tr} [\mathbf{G}_{\mathbf{p_1}} ^A  j_x (\mathbf{p_1}) \mathbf{G}_{\mathbf{p_1}} ^R \mathbf{G}_{\mathbf{p_2}} ^R  j_y (\mathbf{p_2}) \mathbf{G}_{\mathbf{p_2}} ^A 
 \mathbf{G}_{\mathbf{p_3}} ^A ]\Bigg). \notag
\end{align}
This expression contains $\mathbf{p_1}$ and $\mathbf{p_2}$ integrals analogously to the vertex correction, Eq.~\eqref{eq: vc}, there
\[
\sigma_{xy}^{\text{MC}} =0, \; \text{for $n \neq 1$}.
\]
This cancellation originates from the structure of the angular integrals and is not specific to the third moment: higher-order disorder correlators lead to analogous cancellations. Thus, all non-Gaussian disorder moments beyond the second do not generate a finite contribution to the anomalous Hall conductivity in multilayers.

\section{Warped model}
\label{ap: warped}
\subsection{Necessary designations}
\label{sec: necdesignations}
In this section, we introduce all the necessary definitions which we use in this paper for the warped model, see Eqs.\eqref{eq: fermimom1}-\eqref{eq: gammas}.

 Firstly, let us define the Fermi momentum. Using the notation $\epsilon^2- m^2 = v^2 p_0^{2n}$ we find
\begin{align}
    \xi^2 = m^2 + v^2 p^{2n}  + 2 v w p^{2n -3} \cos(3\phi)+ w^2  p^{2n-6}.
\end{align} 
Solving $\xi = \epsilon$ using $\bar p = p/p_0, \tilde w = w/v p_0^3$ leads to the condition
\begin{equation}
    1 = \bar p^{2n}  + \tilde w \bar p^{2n -3} \cos(3\phi)+\tilde  w^2 \bar p^{2n-6}.
    \label{eq: fermimomwarp}
\end{equation}
We cannot solve this equation exactly, but we can find an approximate expression for the $\phi$ dependent Fermi momentum $p_F(\phi) \equiv p_\phi$ in powers of $\tilde w$, Eq.~\eqref{eq: fermimom1}. We remark that $\tilde w$ is a function of ($\epsilon, m$): 
\begin{align}
    &\tilde w = \frac{w}{v^{\frac{n-3}{n}} (\epsilon^2 - m^2)^{3/2n}}=\frac{\gamma_1^{\frac{3}{n}-1} \gamma_2}{m^{3/n}}\cdot \frac{1}{(\frac{\epsilon^2}{m^2} -1)^{3/2n}}.
\end{align}

In the presence of warping we need a $\phi$ dependent replacement for the DOS
\begin{align}
    \rho_{n,\phi}(\epsilon) = \int_0^\infty  \frac{dp p}{2\pi}  \delta(\epsilon - \vert \vec d(p, \phi) \vert),
\end{align}
which leads to Eqs.~\eqref{eq: DOSw}-\eqref{eq: Ff} of the main text.

\subsection{Trilayer graphene in warped model}
In this section, we provide calculations for trilayer graphene, i.e.~$n=3$ in warped case, for Eqs.~\eqref{eq: warped_int_3}, \eqref{eq: correction_warped}, \eqref{eq: warped_nc_3}, \eqref{eq: warped_dif} from the main text.
We start from the effective Hamiltonian $H_3 ^{(w)}$:
\begin{equation}
    H_3 ^{(w)} = \vec d(\v p) \cdot \vec \sigma = \left (\begin{array}{cc}
       m & d_- \\
       d_+ &-m
   \end{array} \right), 
\end{equation}
with $d_-  = d_{x-iy}(\v p) = v \mathcal P ^3 + w; \; d_z  = m$.
\subsubsection{Intrinsic contribution}
\label{sec: intn3w}
Firstly, we consider intrinsic contribution:
\begin{gather}
\sigma_{xy} ^{\text{w,int}}\vert^{n=3} = \frac{e^2}{h} \int \frac{dp^2}{(2\pi)^2} \text{Tr} \Bigg ( j_x (\v p) G_{\v p} ^{R, (w)}  j_y (\v p) G_{\v p} ^{A, (w)} \Bigg), 
\end{gather}

where $G_p ^{R,(w)} = [\epsilon - i0 -H_3 ^{(w)}]^{-1}$. After taking the trace and with using $C_3$ symmetry we deduce:
\begin{align}
    \sigma_{xy} ^{\text{w, int}} \vert^{n=3} & =-\frac{e^2}{h}  \frac{m}{ \epsilon} \frac{3}{2 } \mathcal J_3^{(4)}.
\end{align}
\subsubsection{Vertex correction}
\label{sec: vert3w}

Before considering the non-crossing approximation we should consider vertex correction.
It is a bit more convenient to use bare vertices
\begin{equation}
    J_\pm  = \frac{1}{2} (\partial_{p_x} H \pm i \partial_{p_y} H) \equiv \partial_\pm H,
\end{equation}
which are
\begin{align}
    J_+ & =  \left (\begin{array}{cc}
        0 & \underbrace{3 v \mathcal P^{2} }_{j_+} \\
        0 & 0
    \end{array} \right) = \partial_+ d_- \sigma_+; \;
    J_-  = \left (\begin{array}{cc}
        0 & 0 \\
        \underbrace{3 v \bar{\mathcal P}^{2} }_{j_-} & 0
    \end{array} \right)= \partial_- d_+ \sigma_-,
\end{align}
where  $\sigma_\pm = [\sigma_x \pm i \sigma_y]/2$.

We will need 
\begin{equation}
    \mathcal G^R_{\v p}  =  \frac{1}{\tilde \epsilon^2 - \tilde m^2 - \tilde d_x^2 - \tilde d_y^2},
\end{equation}
with
\begin{align}
    &\tilde \epsilon  = \epsilon + i
    \Gamma; \; \tilde m  = m - i
    \Gamma_z ; \; \notag \\
    &\tilde d_x  = d_x - i
    \Gamma_x ; \; \tilde d_y  = d_y - i
    \Gamma_y \notag.
\end{align}

Partial fraction decomposition leads to
\begin{align}
    &\mathcal G^R_{\v p} \mathcal G^A _{\v p} = - \frac{\text{Im} \mathcal G^R_{\v p}}{2 (\Gamma \epsilon + \Gamma_1 d_x +  \Gamma_2 d_y +\Gamma_3 m)}   \simeq \\
    &\frac{\delta(p-p_{\phi})}{2 \alpha f_{\phi}(p_{\phi})[ \mathcal{I}_n^{(0)} (\epsilon^2 + m^2) + v p_0^3 d_x (\mathcal{I}_n ^{(3)} + \tilde w \mathcal{J}_n ^{(0)})]}.
\end{align}

We find the following first order correction to the vertex
\begin{align}
    \delta J_{\pm} &= \frac{\epsilon}{3 \rho_3(\epsilon)} \alpha \int (dp) j_\pm \mathcal G^R_{\v p} \mathcal G^A _{\v p} N_\pm(\v p), 
\end{align}
with
\begin{align} 
    &N_\pm(\v p) = (d_\pm^2 +(\Gamma_x \pm i \Gamma_y)^2) \sigma_\mp + [ (\epsilon \mp i \Gamma_z)^2-(m\pm i \Gamma)^2] \sigma_\pm + \notag\\
    & d_\pm (m \pm i \Gamma) \sigma_z \pm(i\Gamma_x \mp \Gamma_y)(\epsilon \mp i \Gamma_z) \sigma_z + \notag \\
    & {d_\pm (\epsilon \mp i \Gamma_z)\sigma_0 \pm (i\Gamma_x \mp \Gamma_y)(m \pm i \Gamma)\sigma_0}. \notag
\end{align}
Only terms which have the angular dependence $e^{i 3l \phi}$ with $l \in \mathbb Z$ give the non-zero contribution. With keeping in mind that $d_\pm \sim e^{\pm i 3 \phi}, 1$; $d_\pm^2 \sim e^{\pm i 6 \phi}, e^{\pm i 3\phi}, 1$; $j_\pm \sim e^{\mp i 2 \phi}$ we conclude that vertex correction for $n=3$ is zero.
\subsubsection{Non-crossing approximation}
\label{sec: non3w}

For the non-crossing approximation we should consider the following integral:
\begin{align*}
    &\sigma_{xy}^{\text{w, nc}}\vert^{n=3} = -i \frac{e^2}{h} \int \frac{d^2 p}{(2\pi)^2} \text{Tr}[  j_- \mathbf G^R  j_+ \mathbf G^A -  j_+ \mathbf G^R  j_- \mathbf G^A]=\\
    &\frac{e^2}{h} \int_0^\infty \frac{pdp}{2\pi} \int_0^{2\pi} \frac{d\phi}{2\pi} 36 i (i0) v^2 p^4 \epsilon m \frac{\pi}{2 (0) \epsilon^2} \frac{\delta(p-p_{\phi})}{6 f_{\phi} (p_{\phi})\vert^{n=3}} = \notag \\
     & - \frac{e^2}{h} \frac{3 \epsilon m \mathcal{I}_3^{(0)}}{\epsilon^2 - m^2}  \int_0^{2 \pi} \frac{d \phi}{2 \pi} \frac{   \bar p_{\phi}^5}{ f_{\phi} (\bar p_{\phi})}\times \\
     &\frac{1}{\mathcal{I}_3^{(0)} \frac{\epsilon^2 + m^2}{\epsilon^2 - m^2} + (\bar p_{\phi}^3 \cos{(3 \phi)}  + \tilde w)(\mathcal{I}_3^{(3)} + \tilde w \mathcal{J}_3^{(0)})}.
\end{align*}
We can see if $\tilde w =0$ we get
$
- \frac{e^2}{h} \frac{3 \epsilon m \mathcal{I}_3^{(0)}}{\epsilon^2 - m^2} \frac{1}{\mathcal{I}_3^{(0)} \frac{\epsilon^2 + m^2}{\epsilon^2 - m^2}} = -\frac{e^2}{h} \frac{3 \epsilon m }{\epsilon^2 + m^2}$, i.e.
the answer without warping. 

\subsubsection{X diagram}
\label{sec: X3w}

As in the isotropic model, the X conductivity bubble correspond to:
\begin{widetext}
    \begin{align}
    &\sigma_{xy}^\text{X} = \frac{e^2}{h} \Bigg(\frac{\epsilon }{n \rho_n(\epsilon)} \alpha \Bigg)^2 \int_{\{\mathbf{p}_i\}} \text{Tr} [\mathbf{G}_{\mathbf{p_1}} ^A  j_x (\mathbf{p_1}) \mathbf{G}_{\mathbf{p_1}} ^R \mathbf{G}_{\mathbf{p_3}} ^R \mathbf{G}_{\mathbf{p_2}} ^R j_y (\mathbf{p_2}) \mathbf{G}_{\mathbf{p_2}} ^A \mathbf{G}_{\mathbf{p_4}} ^A ] =\notag \\
    & -\frac{e^2}{h}\Bigg(\frac{\epsilon }{ \rho_n(\epsilon)\pi}\Bigg)^2  \int_{\{\mathbf{p}_i\}}\frac{(2\pi)^2\delta(\mathbf{p_1}+\mathbf{p_2}-\mathbf{p_3}-\mathbf{p_4}) \mathfrak{I}\mathcal{G}_{\mathbf{p_1}}\mathfrak{I}\mathcal{G}_{\mathbf{p_2}} \mathfrak{I}\mathcal{G}_{\mathbf{p_3}} \mathfrak{R}\mathcal{G}_{\mathbf{p_4}} }{[(\epsilon^2 + m^2)\mathcal{I}_3^{(0)} + d_x(\mathbf{p}_1) v p_0^3 (\mathcal{I}_3^{(3)} + \tilde w \mathcal{J}_3^{(0)})] [(\epsilon^2 + m^2)\mathcal{I}_3^{(0)} + d_x(\mathbf{p}_2) v p_0^3 (\mathcal{I}_3^{(3)} + \tilde w \mathcal{J}_3^{(0)})]} \notag \\
    &\Bigg \{ {j_+}(\mathbf{p_2}){j_-}(\mathbf{p_1}) \bigg ( \text{Tr} [N_- (\mathbf{p_1}) (\epsilon + H)_{\mathbf{p_3}} N_+(\mathbf{p_2})  (\epsilon + H)_{\mathbf{p_4}} ] - \text{Tr} [N_- (\mathbf{p_1}) (\epsilon + H)_{\mathbf{p_4}} N_+(\mathbf{p_2})  (\epsilon + H)_{\mathbf{p_3}} ] \bigg ) - \notag \\
    & {j_-}(\mathbf{p_2}){j_+}(\mathbf{p_1}) \bigg ( \text{Tr} [N_- (\mathbf{p_2}) (\epsilon + H)_{\mathbf{p_4}} N_+(\mathbf{p_1})  (\epsilon + H)_{\mathbf{p_3}} ] - \text{Tr} [N_- (\mathbf{p_2}) (\epsilon + H)_{\mathbf{p_3}} N_+(\mathbf{p_1})  (\epsilon + H)_{\mathbf{p_4}} ] \bigg )\Bigg \},
\end{align}
\end{widetext}

where 
\begin{align}
     &N_\pm(\v p) = d_\pm^2 \sigma_\mp +  (\epsilon^2 -m^2) \sigma_\pm + d_\pm m  \sigma_z + d_\pm \epsilon \sigma_0.
\end{align}
After taking momentum integrals, we obtain:
\[
\sigma_{xy}^{\text{X}}\vert^{n=3} = -\frac{e^2}{h} \frac{\epsilon m }{\epsilon^2 - m^2} N_{X,3},
\]
where
\begin{widetext}
\begin{align}
    &N_{X,3}=\frac{1}{6} \int  \frac{d \phi_1}{2\pi}  \frac{d \phi_2}{2\pi}  \frac{d \phi_3}{2\pi} \frac{\bar p_{\phi_1} \bar p_{\phi_2} \bar p_{\phi_3}}{ f_{\phi_1}(\bar p_{\phi_1}) f_{\phi_2}(\bar p_{\phi_2}) f_{\phi_3}(\bar p_{\phi_3})} \frac{ j_+(\bar p_{\phi_2}) j_-(\bar p_{\phi_1})}{1 - d_-(\bar p_{\phi_1}, \bar p_{\phi_2}, \bar p_{\phi_3}) d_+(\bar p_{\phi_1}, \bar p_{\phi_2}, \bar p_{\phi_3})}  \cdot  \notag\\
    & \frac{[d_-(\bar p_{\phi_3}) d_+ (\bar p_{\phi_2}) - d_-(\bar p_{\phi_1}, \bar p_{\phi_2}, \bar p_{\phi_3}) d_+(\bar p_{\phi_2}) + d_-(\bar p_{\phi_1}) d_+ (\bar p_{\phi_3}) - d_-(\bar p_{\phi_1}) d_ +(\bar p_{\phi_1}, \bar p_{\phi_2}, \bar p_{\phi_3})] \cdot  (d_-(\bar p_{\phi_1}) d_+(\bar p_{\phi_2})-1)}{[\frac{(\epsilon^2 + m^2)}{v^2 p_0 ^6 }\mathcal{I}_3^{(0)} + d_x(\bar p_{\phi_1}) (\mathcal{I}_3^{(3)} + \tilde w \mathcal{J}_3^{(0)})] [\frac{(\epsilon^2 + m^2)}{v^2 p_0^6}\mathcal{I}_3^{(0)} + d_x(\bar p_{\phi_2}) (\mathcal{I}_3^{(3)} + \tilde w \mathcal{J}_3^{(0)})]}, 
    \label{eq: NX3}
\end{align} 
\end{widetext}
with
\begin{align*}
    &d_\pm(\bar p_{\phi_1}, \bar p_{\phi_2}, \bar p_{\phi_3}) = (\bar p_{\phi_1} e^{\pm i \phi_1}+\bar p_{\phi_2} e^{\pm i \phi_2}-\bar p_{\phi_3} e^{\pm i \phi_3})^3 + \tilde w
\end{align*}

\subsubsection{$\Psi$ diagram}
\label{sec: Psi3w}
\begin{figure}
    \centering
\includegraphics[width=1\linewidth]{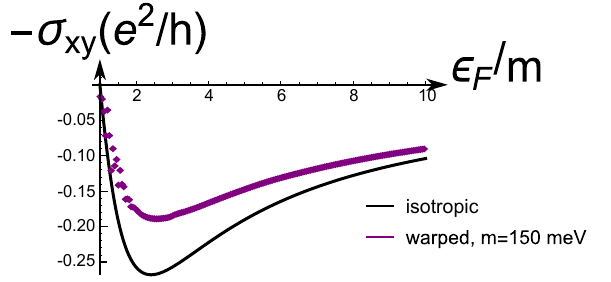}
    \caption{\justifying  Diffractive skew-scattering
component of anomalous Hall conductivity for trilayer graphene.}
    \label{fig: diffractive comparison}
\end{figure}

The contribution from $\Psi$ diagram can be written as
\begin{widetext}
\begin{align}
    &\sigma_{\text{xy}}^{\Psi} =
    -\frac{e^2}{h}\Bigg(\frac{\epsilon }{ \rho_n(\epsilon) \pi } \Bigg)^2 \int_{\{\mathbf{p}_i\}}\frac{(2\pi)^2\delta(\mathbf{p_1 -p_2 -p_3 + p_4}) \mathfrak{I}\mathcal{G}_{\mathbf{p_1}}\mathfrak{I}\mathcal{G}_{\mathbf{p_2}} (\mathfrak{I}\mathcal{G}_{\mathbf{p_4}} \mathfrak{R}\mathcal{G}_{\mathbf{p_3}}+\mathfrak{I}\mathcal{G}_{\mathbf{p_3}} \mathfrak{R}\mathcal{G}_{\mathbf{p_4}})  }{[(\epsilon^2 + m^2)\mathcal{I}_3^{(0)} + d_x(\mathbf{p}_1) v p_0^3 (\mathcal{I}_3^{(3)} + \tilde w \mathcal{J}_3^{(0)})] [(\epsilon^2 + m^2)\mathcal{I}_3^{(0)} + d_x(\mathbf{p}_2) v p_0^3 (\mathcal{I}_3^{(3)} + \tilde w \mathcal{J}_3^{(0)})]} \times \notag \\
    & \Big \{{j_-(\mathbf{p_1})}{j_+(\mathbf{p_2})} \cdot \text{Tr} [N_- (\mathbf{p_1}) (\epsilon + H)_{\mathbf{p_3}} (\epsilon + H)_{\mathbf{p_4}} N_+(\mathbf{p_2}) - N_- (\mathbf{p_1}) N_+(\mathbf{p_2}) (\epsilon + H)_{\mathbf{p_4}} (\epsilon + H)_{\mathbf{p_3}}  ] - \notag \\
    &{j_+(\mathbf{p_1})}{j_-(\mathbf{p_2})} \cdot \text{Tr} [N_+ (\mathbf{p_1}) (\epsilon + H)_{\mathbf{p_3}} (\epsilon + H)_{\mathbf{p_4}} N_-(\mathbf{p_2}) - N_+ (\mathbf{p_1}) N_-(\mathbf{p_2}) (\epsilon + H)_{\mathbf{p_4}} (\epsilon + H)_{\mathbf{p_3}}  ] \Big \}. 
\end{align}
\end{widetext}
After the same steps as for X diagram, we receive:
\begin{align}
\sigma_{xy}^{\Psi}\vert^{n=3} = -\frac{e^2}{h}  \frac{\epsilon m }{(\epsilon^2 - m^2)} \cdot N_{\Psi,3}, 
\end{align}
where
\begin{widetext}
\begin{align}
    &N_{\Psi,3} = \frac{1}{3} \int  \frac{d \phi_1}{2\pi}  \frac{d \phi_2}{2\pi}  \frac{d \phi_3}{2\pi} \frac{\bar p_{\phi_1} \bar p_{\phi_2} \bar p_{\phi_3}}{ f_{\phi_1}(\bar p_{\phi_1}) f_{\phi_2}(\bar p_{\phi_2}) f_{\phi_3}(\bar p_{\phi_3})} \frac{ j_+(\bar p_{\phi_2}) j_-(\bar p_{\phi_1})}{1 - d_-(\bar p_{\phi_1}, \bar p_{\phi_2}, \bar p_{\phi_3}) d_+(\bar p_{\phi_1}, \bar p_{\phi_2}, \bar p_{\phi_3})}  \cdot  \notag \\
    &\frac{  1}{[\frac{(\epsilon^2 + m^2)}{\epsilon^2 - m^2}\mathcal{I}_3^{(0)} + d_x(\bar p_{\phi_1}) (\mathcal{I}_3^{(3)} + \tilde w \mathcal{J}_3^{(0)})] [\frac{(\epsilon^2 + m^2)}{v^2 p_0 ^6}\mathcal{I}_3^{(0)} + d_x(\bar p_{\phi_2})  (\mathcal{I}_3^{(3)} + \tilde w \mathcal{J}_3^{(0)})]} \times \notag \\
    &    \Big [d_-(\bar p_{\phi_1}) d_-(\bar p_{\phi_1}, \bar p_{\phi_2}, \bar p_{\phi_3}) d_+(\bar p_{\phi_2}) \cdot (d_+(\bar p_{\phi_2}) - d_+(\bar p_{\phi_3})) +d_-(\bar p_{\phi_1})^2 d_+(\bar p_{\phi_2}) \cdot (d_+(\bar p_{\phi_3}) -d_+(\bar p_{\phi_2})) +\notag \\
    &  
   d_-(\bar p_{\phi_1}) d_+(\bar p_{\phi_1}, \bar p_{\phi_2}, \bar p_{\phi_3}) \cdot  (d_-(\bar p_{\phi_3}) d_+(\bar p_{\phi_2}) -1) - (d_-(\bar p_{\phi_3}) d_+(\bar p_{\phi_2}) -1)\Big].
   \label{eq: NPsi3}
\end{align}
\end{widetext}
The influence of warping is most clearly manifested in the diffractive skew scattering term. Therefore, we compare the results for this contribution with and without the warping effect in Fig.~\ref{fig: diffractive comparison}.

This concludes the derivation of Eqs.~\eqref{eq: wx3}, \eqref{eq: wpsi3} of the main text. 

\subsection{Tetralayer graphene in warped model}
In this section we provide calculations for tetralayer graphene, i.e.~$n=4$ in the warped case, i.e.~Eqs.~\eqref{eq: warped_int_4}, \eqref{eq: correction_warped}, \eqref{eq: warped_nc_4}, \eqref{eq: warped_dif} from the main text.
We do the similar steps as for the trilayer case.
We start from the effective Hamiltonian $H_4 ^{(w)}$:
\begin{equation}
    H_4 ^{(w)} = \vec d(\v p) \cdot \vec \sigma = \left (\begin{array}{cc}
       m & d_- \\
       d_+ &-m
   \end{array} \right), 
\end{equation}
with $d_-=d_+^\dagger  = d_{x-iy}(\v p) = v \mathcal P ^4 + w\mathcal P; \; d_z  = m.$
\subsubsection{Intrinsic contribution}
\label{sec: intn4w}

Initially, we consider an intrinsic contribution. Doing the same calculations as for trilayer case, see Appendix~\ref{sec: intn3w}, we obtain:
\begin{align}
    &\sigma_{xy} ^{\text{w, int}} \vert^{n=4} =-\frac{e^2}{h}  \frac{m}{ \epsilon} \frac{1}{16 \pi} \int_0^{2\pi} d\phi \frac{\bar p_{\phi} }{f_{\phi} (\bar p_{\phi})}  \times  \notag \\
    &  (16 \bar p_{\phi}^6 + 4 \tilde w \bar p_{\phi}^3 e^{3 i \phi} + 4 \tilde w \bar p_{\phi}^3 e^{-3 i \phi} +\tilde w^2) = \notag \\
    &-\frac{e^2}{h}  \frac{m}{ \epsilon}  \cdot \left(2 \mathcal J_4^{(6)} +\tilde w \mathcal I_4^{(3)} + \frac{\tilde w^2}{8} \mathcal I_4^{(0)}  \right ).
\end{align}

\subsubsection{Vertex correction}
\label{sec: vert4w}
For non-crossing approximation, we should deal with the vertex correction. For tetralayer graphene we have different definitions for currents:
\begin{align}
    &J_+  =  \left (\begin{array}{cc}
        0 & \underbrace{n v \mathcal P^{n-1} + (n-3) w \mathcal P^{n-4}}_{j_+} \\
        0 & 0
    \end{array} \right) = \partial_+ d_- \sigma_+, \notag \\
    &J_-  = \left (\begin{array}{cc}
        0 & 0 \\
        \underbrace{n v \bar{\mathcal P}^{n-1} + (n-3) w \bar{\mathcal P}^{n-4}}_{j_-} & 0
    \end{array} \right)= \partial_- d_+ \sigma_-. \notag
\end{align}
We find the following first order correction to the vertex
\begin{align}
    &\delta J_{\pm} = \frac{\epsilon}{n \rho_n(\epsilon)} \alpha \int (dp) j_\pm \mathcal G^R_{\v p} \mathcal G^A_{\v p} N_\pm(\v p), \notag \\
    &N_\pm(\v p) = d_\pm^2 \sigma_\mp + [ (\epsilon \mp i \Gamma_3)^2-(m\pm i \Gamma)^2] \sigma_\pm + \notag\\
    &d_\pm (m \pm i \Gamma) \sigma_z +d_\pm \epsilon \sigma_0.  
\end{align}
For $n = 4$ we may thus only keep 
    \begin{align}
        &j_\pm N_\pm \doteq j_\pm [ (\epsilon \mp i \Gamma_3)^2-(m\pm i \Gamma)^2] \sigma_\pm \simeq  \\
        &j_\pm [\epsilon^2 -m^2 \mp 2 i (\epsilon \Gamma_3 + m \Gamma)]  = j_\pm [\epsilon^2 -m^2 \mp 2 i \epsilon m  \frac{\pi \alpha}{n} \mathcal I_n^{(0)}]. \notag
    \end{align}
    This leads to 
    \begin{align}
        &\delta J_\pm  = \frac{\sigma_\pm [\epsilon^2 -m^2 \mp 2 i \epsilon m  \frac{\pi \alpha}{n} \mathcal I_n^{(0)}]}{2  (\epsilon^2 + m^2) \rho_n \mathcal I_n^{(0)}}  \times \int (dp) j_\pm \delta(\epsilon - \xi)  = \notag\\
        &\sigma_\pm [\frac{\epsilon^2 - m^2}{2(\epsilon^2 + m^2)} \mp  i \frac{\epsilon m}{\epsilon^2 + m^2}  \frac{\pi \alpha}{n} \mathcal I_n^{(0)}]  \times \left [n vp_0^{n-1} \frac{\mathcal I_n^{(3)}}{\mathcal I_n^{(0)}} +  w \right ].\notag
    \end{align}

 For $n = 4$ we can use that $\delta J_\pm  \propto \sigma_\pm$, so the matrix structure is unchanged. It will be readily clear that this matrix structure is also preserved at higher orders. We can just generally write $ \mathbf J_\pm = \mathbf{j}_\pm \sigma_\pm$
\begin{equation*}
   \mathbf{j}_\pm = \sum_{k = 0}^\infty j_\pm^{(k)},
\end{equation*}
with $j_\pm^{(0)} = j_\pm, j_\pm^{(1)} = \delta j_\pm$. 
We see that for $k \geq 2$
\begin{align}
    j_\pm^{(k)} & = j_\pm^{(k-1)}  \frac{\epsilon}{n \rho_n(\epsilon)} \alpha \int (dp)  \mathcal G^R_{\v p} \mathcal G^A_{\v p} N_\pm(\v p).
\end{align}
Using the previously derived expression for $N_\pm(\v p)$ we see that only the terms $\sigma_\pm$ survive angular integrals. This leads to
\begin{align}
    &j_\pm^{(k)}  = j_\pm^{(k-1)}  \frac{\sigma_\pm [\epsilon^2 -m^2 \mp 2 i \epsilon m  \frac{\pi \alpha}{n} \mathcal I_n^{(0)}]}{2  (\epsilon^2 + m^2) \rho_n \mathcal I_n^{(0)}}  \times \int (dp) \delta(\epsilon - \xi) =\notag \\
        &  j_\pm^{(k-1)} [\underbrace{\frac{\epsilon^2 - m^2}{2(\epsilon^2 + m^2)}}_{=A} \pm  i \underbrace{\frac{-\epsilon m}{\epsilon^2 + m^2}  \frac{\pi \alpha}{n} \mathcal I_n^{(0)}}_{= B}]  = \notag \\
        &\left [4 vp_0^{3} \frac{\mathcal I_4^{(3)}}{\mathcal I_4^{(0)}} +  w \right ] [A \pm i B]^{k}.
\end{align}
We thus find in general that 
\begin{align}
    &\mathbf j_\pm  \simeq 
    {[4 v p_0^3 \bar p^3 e^{\mp i 3 \phi} + w]}+ {\left [4 vp_0^{3} \frac{\mathcal I_4^{(3)}}{\mathcal I_4^{(0)}} +  w \right ]  \frac{A}{1-A}} \pm \notag \\
    &i {\left [4 vp_0^{3} \frac{\mathcal I_4^{(3)}}{\mathcal I_4^{(0)}} +  w \right ]   \frac{ B}{(1- A)^2}} = j_{\pm} + \delta \mathbf{j'} \pm i \delta \mathbf{j''}.
\end{align}

\subsubsection{Non-crossing approximation}
\label{sec: non4w}

For the non-crossing approximation we should consider the following integral:
\begin{align}
    \sigma_{xy}^{\text{nc}} = -i \frac{e^2}{h} \int \frac{d^2 p}{(2\pi)^2} \text{Tr}[ \mathbf J_- \mathbf G^R  J_+ \mathbf G^A - \mathbf J_+ \mathbf G^R  J_- \mathbf G^A], \notag
\end{align}
where we put all the vertex correction only to the left vertex.
Then
\begin{align}
    &\sigma_{xy}^{\text{w, nc}}\vert^{n=4}  =  -i\frac{e^2 }{h} \int (dp) \mathcal G^R_{\v p} \mathcal G^A_{\v p}  \times \\
    &\{ \mathbf j_- j_+\tr[N_-(\v p)\sigma_+] - \mathbf j_+ j_-\tr[N_+(\v p)\sigma_-] \}= - \frac{e^2}{h} \frac{4 m \epsilon}{\epsilon^2 + m^2} \times \notag \\
    &   \Big \{\left [\frac{\mathcal J_4^{(6)}}{\mathcal I_4^{(0)}} - \frac{[\mathcal I_4^{(3)}]^2}{[\mathcal I_4^{(0)}]^2} \right]   + \frac{4 \left(\epsilon ^2 + m^2\right)^2}{\left(\epsilon ^2 + 3 m^2\right)^2} \left [ \frac{\mathcal I_4^{(3)}}{\mathcal I_4^{(0)}}  + \tilde w/4\right ]^2\Big \}\notag .
\end{align}

\subsubsection{X diagram}
\label{sec: X4w}

We should do the similar calculations as for trilayer case. We just notice that for $n=4$:
\begin{align}
    \mathbf{J_+}(\mathbf{p_2})\mathbf{J_-}(\mathbf{p_1})= j_-(\v p_1) j_+(\v p_2) + \delta \mathbf{j'} (j_-(\v p_1) +j_+(\v p_2)) + \delta \mathbf{j'}^2. \notag
\end{align}
We introduce the following notations for tetralayer graphene:
\begin{align}
    &\delta \bar j' = \left [4 \frac{\mathcal I_4^{(3)}}{\mathcal I_4^{(0)}} +  \tilde w \right ]  \frac{\epsilon^2 - m^2}{\epsilon^2 + 3 m^2} ; \; \notag\\
    &d_{\pm}(\bar p_{\phi_1}, \bar p_{\phi_2}, \bar p_{\phi_3}) = (\bar p_{\phi_1} e^{\pm i \phi_1}+\bar p_{\phi_2} e^{\pm i \phi_2}-\bar p_{\phi_3} e^{\pm i \phi_3})^4 + \notag\\
    &\tilde w (\bar p_{\phi_1} e^{\pm i \phi_1}+\bar p_{\phi_2} e^{\pm i \phi_2}-\bar p_{\phi_3} e^{\pm i \phi_3}).
\end{align}
And we can get 
\begin{align}
&\sigma_{xy}^{\text{X}}\vert^{n=4}= -\frac{e^2}{h}  \frac{\epsilon m (\epsilon^2 -m^2)}{(\epsilon^2 + m^2)^2} \cdot N_{X,4}\vert^{1)}+ \notag \\
&-\frac{e^2}{h}  \frac{\epsilon m (\epsilon^2 - m^2)^2}{(\epsilon^2 + 3 m^2)(\epsilon^2+m^2)^2} \cdot N_{X,4}\vert^{2)}+\notag \\
&-\frac{e^2}{h}  \frac{\epsilon m (\epsilon^2 - m^2)^3}{(\epsilon^2 + 3 m^2)^2(\epsilon^2+m^2)^2} \cdot N_{X,4}\vert^{3)},
\end{align} 

where
\begin{widetext}
\begin{align}
    &N_{X,4}\vert^{1)} = \frac{1}{8 \left(\mathcal {I}_4^{(0)}\right)^2} \int  \frac{d \phi_1}{2\pi}  \frac{d \phi_2}{2\pi}  \frac{d \phi_3}{2\pi} \frac{\bar p_{\phi_1} \bar p_{\phi_2} \bar p_{\phi_3}}{ f_{\phi_1}(\bar p_{\phi_1}) f_{\phi_2}(\bar p_{\phi_2}) f_{\phi_3}(\bar p_{\phi_3})}\frac{j_+(\bar p_{\phi_2})j_-(\bar p_{\phi_1})}{1- d_-(\bar p_{\phi_1}, \bar p_{\phi_2}, \bar p_{\phi_3})d_+(\bar p_{\phi_1},\bar p_{\phi_2}, \bar p_{\phi_3}) }  \cdot \notag \\
    &(d_-(\bar p_{\phi_1}) d_+(\bar p_{\phi_2}) -1) \cdot [d_-(\bar p_{\phi_3}) d_+ (\bar p_{\phi_2}) - d_-(\bar p_{\phi_1}, \bar p_{\phi_2}, \bar p_{\phi_3}) d_+(\bar p_{\phi_2}) + d_-(\bar p_{\phi_1}) d_+ (\bar p_{\phi_3}) - d_-(\bar p_{\phi_1}) d_ +(\bar p_{\phi_1}, \bar p_{\phi_2}, \bar p_{\phi_3})],
    \label{eq: NX41}
\end{align}
\begin{align}
    &N_{X,4}\vert^{2)} = \frac{1}{8 \left(\mathcal {I}_4^{(0)}\right)^2} \cdot \left [4 \frac{\mathcal I_4^{(3)}}{\mathcal I_4^{(0)}} +  \tilde w \right ] \int  \frac{d \phi_1}{2\pi}  \frac{d \phi_2}{2\pi}  \frac{d \phi_3}{2\pi} \frac{\bar p_{\phi_1} \bar p_{\phi_2} \bar p_{\phi_3}}{ f_{\phi_1}(\bar p_{\phi_1}) f_{\phi_2}(\bar p_{\phi_2}) f_{\phi_3}(\bar p_{\phi_3})}\frac{j_+(\bar p_{\phi_2}) + j_-(\bar p_{\phi_1}) }{1- d_-(\bar p_{\phi_1}, \bar p_{\phi_2}, \bar p_{\phi_3})d_+(\bar p_{\phi_1},\bar p_{\phi_2}, \bar p_{\phi_3}) }  \cdot \notag \\
    &(d_-(\bar p_{\phi_1}) d_+(\bar p_{\phi_2}) -1) \cdot [d_-(\bar p_{\phi_3}) d_+ (\bar p_{\phi_2}) - d_-(\bar p_{\phi_1}, \bar p_{\phi_2}, \bar p_{\phi_3}) d_+(\bar p_{\phi_2}) + d_-(\bar p_{\phi_1}) d_+ (\bar p_{\phi_3}) - d_-(\bar p_{\phi_1}) d_ +(\bar p_{\phi_1}, \bar p_{\phi_2}, \bar p_{\phi_3})],
    \label{eq: NX42}
\end{align}
\begin{align}
    &N_{X,4}\vert^{3)} = \frac{1}{8 \left(\mathcal {I}_4^{(0)}\right)^2} \cdot \left [4 \frac{\mathcal I_4^{(3)}}{\mathcal I_4^{(0)}} +  \tilde w \right ]^2 \int  \frac{d \phi_1}{2\pi}  \frac{d \phi_2}{2\pi}  \frac{d \phi_3}{2\pi} \frac{\bar p_{\phi_1} \bar p_{\phi_2} \bar p_{\phi_3}}{ f_{\phi_1}(\bar p_{\phi_1}) f_{\phi_2}(\bar p_{\phi_2}) f_{\phi_3}(\bar p_{\phi_3})}\frac{1}{1- d_-(\bar p_{\phi_1}, \bar p_{\phi_2}, \bar p_{\phi_3})d_+(\bar p_{\phi_1},\bar p_{\phi_2}, \bar p_{\phi_3}) }  \cdot \notag\\
    &(d_-(\bar p_{\phi_1}) d_+(\bar p_{\phi_2}) -1) \cdot [d_-(\bar p_{\phi_3}) d_+ (\bar p_{\phi_2}) - d_-(\bar p_{\phi_1}, \bar p_{\phi_2}, \bar p_{\phi_3}) d_+(\bar p_{\phi_2}) + d_-(\bar p_{\phi_1}) d_+ (\bar p_{\phi_3}) - d_-(\bar p_{\phi_1}) d_ +(\bar p_{\phi_1}, \bar p_{\phi_2}, \bar p_{\phi_3})].
    \label{eq: NX43}
\end{align}
\end{widetext}

\subsubsection{$\Psi$ diagram}
\label{sec: Psi4w}

After the same steps as for X-diagram, we can deduce that:
\begin{widetext}
\begin{align}
\sigma_{xy}^{\Psi}\vert^{n=4}=-\frac{e^2}{h}  \frac{\epsilon m (\epsilon^2 -m^2)}{(\epsilon^2 + m^2)^2} \cdot N_{\Psi,4}\vert^{1)} - \frac{e^2}{h}  \frac{\epsilon m (\epsilon^2 - m^2)^2}{(\epsilon^2 + 3 m^2)(\epsilon^2+m^2)^2} \cdot N_{\Psi,4}\vert^{2)} -\frac{e^2}{h}  \frac{\epsilon m (\epsilon^2 - m^2)^3}{(\epsilon^2 + 3 m^2)^2(\epsilon^2+m^2)^2} \cdot N_{\Psi,4}\vert^{3)},
\end{align}
\begin{align}
    &N_{\Psi,4}\vert^{1)} = \frac{1}{4 \left(\mathcal {I}_4^{(0)}\right)^2} \int  \frac{d \phi_1}{2\pi}  \frac{d \phi_2}{2\pi}  \frac{d \phi_3}{2\pi} \frac{\bar p_{\phi_1} \bar p_{\phi_2} \bar p_{\phi_3}}{ f_{\phi_1}(\bar p_{\phi_1}) f_{\phi_2}(\bar p_{\phi_2}) f_{\phi_3}(\bar p_{\phi_3})} \frac{1}{1- d_-(\bar p_{\phi_1}, \bar p_{\phi_2}, \bar p_{\phi_3})d_+(\bar p_{\phi_1},\bar p_{\phi_2}, \bar p_{\phi_3}) }\notag \\
    & j_+(\bar p_{\phi_2})j_-(\bar p_{\phi_1}) \cdot \Big [d_-(\bar p_{\phi_1}) d_-(\bar p_{\phi_1}, \bar p_{\phi_2}, \bar p_{\phi_3}) d_+(\bar p_{\phi_2}) \cdot (d_+(\bar p_{\phi_2}) - d_+(\bar p_{\phi_3})) +d_-(\bar p_{\phi_1})^2 d_+(\bar p_{\phi_2}) \cdot (d_+(\bar p_{\phi_3}) -d_+(\bar p_{\phi_2})) +\notag \\
    &  d_-(\bar p_{\phi_1}) d_+(\bar p_{\phi_1}, \bar p_{\phi_2}, \bar p_{\phi_3}) \cdot  (d_-(\bar p_{\phi_3}) d_+(\bar p_{\phi_2}) -1) - (d_-(\bar p_{\phi_3}) d_+(\bar p_{\phi_2}) -1)\Big],
    \label{eq: NPsi41}
\end{align}
\begin{align}
    N_{\Psi,4}\vert^{2)} & =  \frac{1}{4\left(\mathcal {I}_4^{(0)}\right)^2} \cdot \left [4 \frac{\mathcal I_4^{(3)}}{\mathcal I_4^{(0)}} +  \tilde w \right ] \int \frac{d \phi_1}{2\pi}  \frac{d \phi_2}{2\pi}  \frac{d \phi_3}{2\pi} \frac{\bar p_{\phi_1} \bar p_{\phi_2} \bar p_{\phi_3}}{ f_{\phi_1}(\bar p_{\phi_1}) f_{\phi_2}(\bar p_{\phi_2}) f_{\phi_3}(\bar p_{\phi_3})} \frac{1}{1- d_-(\bar p_{\phi_1}, \bar p_{\phi_2}, \bar p_{\phi_3})d_+(\bar p_{\phi_1},\bar p_{\phi_2}, \bar p_{\phi_3}) }\notag \\
    & (j_+(\bar p_{\phi_2})+j_-(\bar p_{\phi_1})) \cdot \Big [d_-(\bar p_{\phi_1}) d_-(\bar p_{\phi_1}, \bar p_{\phi_2}, \bar p_{\phi_3}) d_+(\bar p_{\phi_2}) \cdot (d_+(\bar p_{\phi_2}) - d_+(\bar p_{\phi_3})) +d_-(\bar p_{\phi_1})^2 d_+(\bar p_{\phi_2}) \cdot (d_+(\bar p_{\phi_3}) -d_+(\bar p_{\phi_2})) +\notag \\
    &  
   d_-(\bar p_{\phi_1}) d_+(\bar p_{\phi_1}, \bar p_{\phi_2}, \bar p_{\phi_3}) \cdot  (d_-(\bar p_{\phi_3}) d_+(\bar p_{\phi_2}) -1) - (d_-(\bar p_{\phi_3}) d_+(\bar p_{\phi_2}) -1)\Big],
   \label{eq: NPsi42}
\end{align}
\begin{align}
    &N_{\Psi,4}\vert^{3)} = \frac{1}{4 \left(\mathcal {I}_4^{(0)}\right)^2} \cdot \left [4 \frac{\mathcal I_4^{(3)}}{\mathcal I_4^{(0)}} +  \tilde w \right ]^2 \int \frac{d \phi_1}{2\pi}  \frac{d \phi_2}{2\pi}  \frac{d \phi_3}{2\pi} \frac{\bar p_{\phi_1} \bar p_{\phi_2} \bar p_{\phi_3}}{ f_{\phi_1}(\bar p_{\phi_1}) f_{\phi_2}(\bar p_{\phi_2}) f_{\phi_3}(\bar p_{\phi_3})} \frac{1}{1- d_-(\bar p_{\phi_1}, \bar p_{\phi_2}, \bar p_{\phi_3})d_+(\bar p_{\phi_1},\bar p_{\phi_2}, \bar p_{\phi_3}) }\notag \\
    & \Big [d_-(\bar p_{\phi_1}) d_-(\bar p_{\phi_1}, \bar p_{\phi_2}, \bar p_{\phi_3}) d_+(\bar p_{\phi_2}) \cdot (d_+(\bar p_{\phi_2}) - d_+(\bar p_{\phi_3})) +d_-(\bar p_{\phi_1})^2 d_+(\bar p_{\phi_2}) \cdot (d_+(\bar p_{\phi_3}) -d_+(\bar p_{\phi_2})) + \notag \\
    &  
   d_-(\bar p_{\phi_1}) d_+(\bar p_{\phi_1}, \bar p_{\phi_2}, \bar p_{\phi_3}) \cdot  (d_-(\bar p_{\phi_3}) d_+(\bar p_{\phi_2}) -1) - (d_-(\bar p_{\phi_3}) d_+(\bar p_{\phi_2}) -1)\Big].
   \label{eq: NPsi43}
\end{align}
\end{widetext}

\begin{figure}
    \centering
\includegraphics[width=1\linewidth]{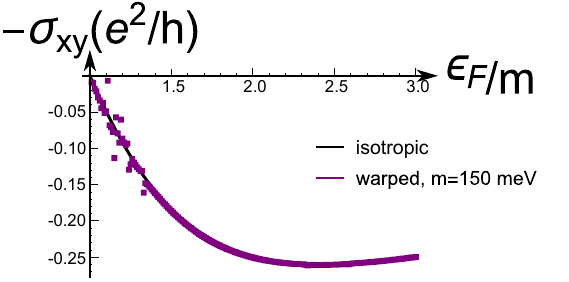}
    \caption{\justifying  Diffractive skew-scattering
component of anomalous Hall conductivity for tetralayer graphene.}
    \label{fig: diffractive comparison4}
\end{figure}

In tetralayer graphene, the influence of warping is generally weak across all contributions. For consistency with the trilayer case, we present in Fig.~\ref{fig: diffractive comparison4} only the results for diffractive skew scattering at the low energy region, where the absolute value of conductivity is decreased; however, even for this mechanism, the effect remains barely discernible.

\end{document}